\documentclass[aps,pra,nofootinbib,twocolumn]{revtex4-2}
\usepackage{eurosym}
\usepackage{amsmath, amsthm, amscd, amssymb}
\usepackage{bm}
\usepackage{bbm}
\usepackage{epsfig}
\usepackage[sort&compress]{natbib}

\begin{document}

\title{Quantum evolution of mixed states and performance of quantum heat
engines}
\author{Anatoly A. Svidzinsky$^{*}$ and Wenzhuo Zhang}
\affiliation{Institute for Quantum Science and Engineering, Department of Physics \& Astronomy, Texas A\&M
University, College Station, TX 77843 \\
$^*$Author to whom any correspondence should be addressed. 
E-mail: asvid@physics.tamu.edu}

\begin{abstract}
We introduce a technique for calculating the density operator time evolution
along the lines of Heisenberg representation of quantum mechanics. Using
this technique, we find the exact solution for the quantum evolution of two and
three coupled harmonic oscillators initially prepared in thermal states at
different temperatures. We show that such systems exhibit interesting
quantum dynamics in which oscillators swap their thermal states due to
correlation induced in the process of energy exchange and yield noise
induced coherence. A photonic quantum heat engine (QHE) composed of two
optical cavities can be modeled as coupled harmonic oscillators with
time-dependent frequencies. Photons in the cavities become correlated during
the engine operation. We show that the work done by such an engine is
maximum if at the end of the cycle the oscillators swap numbers of
excitations which can be achieved when the engine operates under the
condition of parametric resonance. We also show that Carnot formula yields
limiting efficiency for QHEs under general assumptions. Moreover, we show
that, by making a canonical transformation, density operator of arbitrary $n$%
-mode Gaussian state can be written as a product of $n$ thermal density
operators describing independent collective excitations with different
temperatures. Thus, operation of QHEs based on the correlated Gaussian
states is equivalent to that based on uncorrelated thermal reservoirs. Our
results deepen understanding of quantum evolution of mixed states which
could be useful to design quantum machines with better performance.
\end{abstract}

\date{\today }
\maketitle

\section{Introduction}

Quantum evolution of a system can be qualitatively different from the
classical counterpart. For example, the increase of electric current along a
normal metal ring yields heating. However, if the metal ring is in a quantum
superconducting state the adiabatic increase of the electric current can
result in a cooling effect \cite{Svid02}. Optical pumping and laser cooling
of atoms are among other examples of quantum refrigeration.

Heat engines constitute the major building blocks of modern technologies.
Classical heat engines and refrigerators serve as impeccable applications
for centuries. Unconventional properties of the quantum world \cite{Zhan26}
have inspired development of the quantum heat engines (QHEs). Some of them
(lasers, solar cells, etc.) have already received broad applications. Recent
technological advances suggest that we are in the midst of a new revolution
in quantum physics, which is about controlling individual quantum systems
(e.g., atoms and molecules) to a greater extent, enabling more powerful
applications of quantum mechanics \cite{Jaed18}. This may lead to the
development of QHEs at the nanoscales.

The operation mechanism of the QHE is described by the laws of quantum
mechanics: the working fluid is quantum in nature and quantum effects are
not negligible. Whether quantum evolution can lead to a violation of the
second law of thermodynamics is a fundamental question which yet remains
open. One of the formulations of the second law is associated with the
Carnot's theorem \cite{Carn24} which states that the efficiency of any heat
engine that uses hot thermal reservoir at temperature $T_{h}$ as an energy
source and a cold reservoir at temperature $T_{c}$ as an entropy sink cannot
exceed the Carnot efficiency%
\begin{equation*}
\eta _{\text{max}}=1-\frac{T_{c}}{T_{h}}.
\end{equation*}

Whether quantum effects can affect performance of QHEs is a subject of
investigation in recent decades which has led to the creation of a separate
area of study, quantum thermodynamics \cite{Mill16,Bind18}. This is
especially important for small machines consisting of a small number of
particles. Discussions of QHEs spread a wide range (for recent reviews see,
e.g., \cite{Kosl17,Mukh21,Bhat21,Myer22d}).

When the heat baths are not thermal (e.g. there is coherence, squeezing,
quantum correlations, entanglement, etc.) the \textquotedblleft
efficiency\textquotedblright\ of quantum thermal machines could surpass the
thermodynamic Carnot bound \cite%
{Scul01,Scul03,Dill09,Scul10,Huan12,Raha12,Abah14,Ross14,Hard15,Nied16,Manz16,Klae17,Agar17}%
. However, there is no violation of the laws of thermodynamics since extra
resources are required to prepare the baths in non-thermal states which is
not included in the estimate of the engine efficiency. When this energetic
cost is properly accounted for, the Carnot efficiency is maintained as the
fundamental bound on the cycle efficiency \cite{Gard15}.

Operation of a QHE between hot and cold thermal reservoirs could lead to
generation of coherence in the working substance which can improve the
engine's performance. For example, induced coherence can enhance power of a
photocell by increasing absorption of solar radiation \cite%
{Svid11,Scul11,Svid12} and could improve energy transport in photosynthetic
reaction centers \cite{Dorf13}. The latter can be seen as a biological QHE
that transforms high-energy thermal photon radiation into low-entropy
electron flux.

QHEs have been examined in many platforms including implementations in
single spin systems \cite{Feld00,Geva92,Hern07}, coupled spin systems \cite%
{Huan13,Huan14,Feld04}, harmonic oscillators \cite{Abah12,Abah16,Myer20},
relativistic oscillators \cite{Myer21b}, an ideal Bose gas \cite{Wang09}, a
Bose--Einstein condensate \cite{Myer22}, anyons \cite{Myer21,Myer21a}, a
two-level atom \cite{Wang12}, atomic systems with multiple energy levels 
\cite{Quan05,Huan18,Klim18}, optomechanical QHEs with phonon coupling \cite%
{Zhan14,Zhan14a,Zhan17}, engines based on quantum Hall edge states \cite%
{Soth14,Sanc15a,Sanc15}, on electromagnetically induced transparency \cite%
{Harr16}, and so on.

In the past decade, quantum engine cycles were experimentally demonstrated
in various systems such as cold atoms \cite{Zou17,Bout21}, a nanobeam \cite%
{Klae17}, an ensemble of nitrogen-vacancy centres \cite{Klat19}, nuclear
spins \cite{Pete19}, using a trapped ion as working fluid \cite%
{Abah12,Ross14,Ross16}, and using a spin coupled to single-ion motion \cite%
{Lind19,Horn20}. A photonic quantum engine driven by superradiance was
reported in \cite{Kim22}.

Significant progress in the fabrication of mechanical devices at the micro-
and nano scale has been achieved \cite{Mido18}. Hybrid systems that couple
optical, electrical and mechanical degrees of freedom via radiation pressure
in nanoscale devices are under development in laboratories worldwide \cite%
{Aspel14}. Ground-state cooling of mechanical resonators \cite{Teuf11,Pete16}%
, and manipulation of photons and phonons at quantum levels \cite%
{Hong17,Kotl21} have been realized in these systems.

The nano-opto-electro-mechanical systems offer unprecedented opportunities
to control light in nanophotonic structures which provides an excellent
platform for studying quantum thermodynamics and nano QHEs. The optical
cavity enhances the radiation pressure force and provides a feedback
mechanism which allows one to control the motion of the mechanical resonator
at mesoscopic scales, and to investigate the underlying physical mechanisms
under a variety of parameter conditions.

In this paper, we continue the discussion of QHEs by studying specific
examples of quantum evolution for which solutions can be obtained
analytically. We investigate quantum evolution of mixed states using a
formalism similar to the Heisenberg representation of quantum mechanics in
which the operators incorporate a dependency on time. However, in the
present treatment, equation of motion for the operators differs from the
conventional Heisenberg equation by the time reversal transformation $%
t\rightarrow -t$. That is operators obey the same evolution equation as the
density operator. The present formalism is applicable when a mixed state
undergoes a unitary evolution. That is during evolution, the system,
initially prepared in a mixed state, is isolated from the environment.

First (Sects. \ref{Sec2} and \ref{Sec3}), we analyze two and three coupled
harmonic oscillators as well as two-level atoms initially prepared in
thermal states at different temperatures. We show that such systems exhibit
interesting quantum dynamics in which oscillators swap temperatures due to
correlations induced in the process of energy exchange and yield noise
induced coherence. As we show, for this example the state of the system with
coherence can be represented as a product of thermal states with different
temperatures using collective modes as the basis set. As a consequence, one
can extract work from a single heat reservoir with coherence in just the
same way as if QHE operates between hot and cold thermal baths.

In the following Sections \ref{Sec4}-\ref{Sec7}, we investigate heat engines
with quantum working substance that use hot and cold reservoirs in thermal
equilibrium as the energy source and the entropy sink respectively. To be
specific, we consider a working substance that can be modeled as coupled
harmonic oscillators (photonic QHE) or coupled two-level systems (atomic
QHE) whose transition frequencies depend on time. We show that the engine
efficiency depends only on the initial frequencies and is independent of how
oscillator frequencies vary during the engine operation. We also show that
the work done by such engine is maximum if at the end of the cycle the
oscillators swap numbers of excitations which can be achieved when the
engine operates under the condition of parametric resonance. We also discuss
the engine operation protocol which yields maximum engine power.

In Sect. \ref{Sec6}, we find that oscillators coupled only by
counter-rotating terms (BEC-like working substance) are unable to produce
positive work in any closed cycle. Finally (Sect. \ref{Sec7}),\ we show that
Carnot formula yields the limiting efficiency for QHEs under general
assumptions, which provides an alternative way to derive the Carnot limit on
QHE efficiency \cite{Pusz78,Alic79}.

Moreover, we show that arbitrary correlated Gaussian states are equivalent
to a set of independent collective thermal modes at different temperatures
(see Sect. \ref{Gaussian}). Hence, the Carnot formula involving effective
temperatures can be used to find the maximum efficiency of QHEs operating
based on such correlated reservoirs.

\section{Quantum evolution of mixed states}

\label{Sec2}

\subsection{Coupled oscillators in thermal state}

In this section, we introduce a technique which allows us to find the time
evolution of the density operator describing an isolated quantum system
initially prepared in a mixed state. To be specific, we consider two coupled
harmonic oscillators $a$ and $b$ with frequencies $\omega _{a}(t)$ and $%
\omega _{b}(t)$ governed by the Hermitian Hamiltonian 
\begin{equation}
\hat{H}=\hslash \omega _{a}(t)\hat{a}^{\dag }\hat{a}+\hslash \omega _{b}(t)%
\hat{b}^{\dag }\hat{b}+\hslash g\left( \hat{a}^{\dag }\hat{b}+\hat{a}\hat{b}%
^{\dag }\right)  \label{q1}
\end{equation}%
which depends on time through $\omega _{a,b}(t)$. In Eq. (\ref{q1}),
operators $\hat{a}$ and $\hat{b}$ obey bosonic commutation relations. We
assume that initially both oscillators are in thermal states with
temperatures $T_{a}$ and $T_{b}$ respectively, and $T_{a}>T_{b}$. The
initial density operator of the system has a separable form and reads%
\begin{equation}
\hat{\rho}_{0}(\hat{a},\hat{b})=Ne^{-q\hat{a}^{\dag }\hat{a}}e^{-p\hat{b}%
^{\dag }\hat{b}},  \label{q2}
\end{equation}%
where $q=\hslash \omega _{a}/k_{B}T_{a}$, $p=\hslash \omega _{b}/k_{B}T_{b}$
and $N$ is the normalization factor%
\begin{equation*}
N=\left( 1-e^{-q}\right) \left( 1-e^{-p}\right) .
\end{equation*}

Evolution of the density operator of the isolated system is described by the
Liouville--von Neumann equation%
\begin{equation}
i\hslash \frac{\partial \hat{\rho}}{\partial t}=\left[ \hat{H},\hat{\rho}%
\right] ,  \label{q3}
\end{equation}%
where the square brackets stand for the commutator of the operators $\left[ 
\hat{H},\hat{\rho}\right] =\hat{H}\hat{\rho}-\hat{\rho}\hat{H}$. Equation (%
\ref{q3}), subject to the initial condition (\ref{q2}), has the following
formal solution%
\begin{equation}
\hat{\rho}(t)=\hat{U}^{\dag }(t)\hat{\rho}_{0}(\hat{a},\hat{b})\hat{U}(t),
\label{q3a}
\end{equation}%
where $\hat{U}(t)$ is a unitary operator obeying equation 
\begin{equation}
-i\hslash \frac{\partial \hat{U}(t)}{\partial t}=\hat{U}(t)\hat{H},
\end{equation}%
and satisfying the initial condition $\hat{U}(0)=1$. The operator $\hat{U}%
(t) $ differs from the conventional evolution operator by the time reversal $%
t\rightarrow -t$. Using the Taylor expansion of the operator $\hat{\rho}_{0}(%
\hat{a},\hat{b})$, we obtain that solution (\ref{q3a}) can be written as
(see Appendix A) 
\begin{equation}
\hat{\rho}(t)=\hat{\rho}_{0}(\hat{A}(t),\hat{B}(t))=Ne^{-q\hat{A}^{\dag }(t)%
\hat{A}(t)}e^{-p\hat{B}^{\dag }(t)\hat{B}(t)},  \label{q3b}
\end{equation}%
where 
\begin{equation}
\hat{A}(t)=\hat{U}^{\dag }(t)\hat{a}\hat{U}(t),  \label{w1}
\end{equation}%
\begin{equation}
\hat{B}(t)=\hat{U}^{\dag }(t)\hat{b}\hat{U}(t).  \label{w2}
\end{equation}%
Equations (\ref{w1}) and (\ref{w2}) yield that operators $\hat{A}(t)$ and $%
\hat{B}(t)$ obey the same equation of motion (\ref{q3}) as the density
operator $\hat{\rho}(t)$, namely%
\begin{equation}
i\hslash \frac{\partial \hat{A}(t)}{\partial t}=\left[ \hat{H},\hat{A}(t)%
\right] ,  \label{w3}
\end{equation}%
\begin{equation}
i\hslash \frac{\partial \hat{B}(t)}{\partial t}=\left[ \hat{H},\hat{B}(t)%
\right] ,  \label{w3a}
\end{equation}%
and are subject to the initial conditions $\hat{A}(0)=\hat{a}$ and $\hat{B}%
(0)=\hat{b}$. In addition, Eqs. (\ref{w1}) and (\ref{w2}) yield that
operators $\hat{A}(t)$ and $\hat{B}(t)$ obey the same bosonic commutation
relations as the operators $\hat{a}$ and $\hat{b}$, namely 
\begin{equation}
\lbrack \hat{A}(t),\hat{A}^{\dag }(t)]=1,\quad \lbrack \hat{B}(t),\hat{B}%
^{\dag }(t)]=1,
\end{equation}%
and all other commutators, including $[\hat{A}(t),\hat{B}^{\dag }(t)]$, are
equal to zero. We look for solutions of Eqs. (\ref{w3}) and (\ref{w3a}) in
the form%
\begin{equation}
\hat{A}(t)=C(t)\hat{a}+D(t)\hat{b},\quad \hat{B}(t)=E(t)\hat{b}+F(t)\hat{a},
\label{y0}
\end{equation}%
which yields the following equations for the unknown functions $C(t)$, $%
D(t), $ $E(t)$ and $F(t)$ \ 
\begin{equation}
i\dot{C}=-\omega _{a}(t)C-gD,\quad i\dot{D}=-\omega _{b}(t)D-gC,  \label{y1}
\end{equation}%
\begin{equation}
i\dot{E}=-\omega _{b}(t)E-gF,\quad i\dot{F}=-\omega _{a}(t)F-gE,  \label{y2}
\end{equation}%
subject to the initial conditions%
\begin{equation}
C(0)=1,\quad D(0)=0,\quad E(0)=1,\quad F(0)=0.  \label{y2a}
\end{equation}%
Equations (\ref{y1})-(\ref{y2a}) have the following integrals of motion 
\begin{equation}
|C(t)|^{2}+|D(t)|^{2}=1,  \label{p1}
\end{equation}%
\begin{equation}
|E(t)|^{2}+|F(t)|^{2}=1,  \label{p2}
\end{equation}%
\begin{equation}
C(t)F^{\ast }(t)+D(t)E^{\ast }(t)=0,  \label{p3}
\end{equation}%
which can be also obtained directly from the commutation relations for the
operators $\hat{A}(t)$ and $\hat{B}(t)$. For example, $[\hat{A}(t),\hat{A}%
^{\dag }(t)]=1$ yields Eq. (\ref{p1}), etc.

If oscillators have the same frequency $\omega $ which is independent of
time, the solution of Eqs. (\ref{y1})-(\ref{y2a}) describes Rabi
oscillations, namely,%
\begin{equation}
\hat{A}(t)=e^{i\omega t}\left( \cos (gt)\hat{a}+i\sin (gt)\hat{b}\right)
\label{M1}
\end{equation}%
\begin{equation}
\hat{B}(t)=e^{i\omega t}\left( \cos (gt)\hat{b}+i\sin (gt)\hat{a}\right) ,
\label{M2}
\end{equation}%
and the time evolution of the system's density operator is given by the
formula%
\begin{equation*}
\hat{\rho}(t)=Ne^{-q\left( \cos (gt)\hat{a}^{\dag }-i\sin (gt)\hat{b}^{\dag
}\right) \left( \cos (gt)\hat{a}+i\sin (gt)\hat{b}\right) }
\end{equation*}%
\begin{equation}
\times e^{-p\left( \cos (gt)\hat{b}^{\dag }-i\sin (gt)\hat{a}^{\dag }\right)
\left( \cos (gt)\hat{b}+i\sin (gt)\hat{a}\right) }.  \label{p4}
\end{equation}%
Equation (\ref{p4}) shows that $\hat{\rho}(t)$ periodically oscillates with
the Rabi frequency $2g$ and the energy flows back and forth between the
oscillators. For the time dependence of the average oscillator energy we
obtain%
\begin{equation}
E_{a}(t)=\hslash \omega \left( \frac{\cos ^{2}(gt)}{e^{q}-1}+\frac{\sin
^{2}(gt)}{e^{p}-1}\right) ,  \label{EA}
\end{equation}%
\begin{equation}
E_{b}(t)=\hslash \omega \left( \frac{\cos ^{2}(gt)}{e^{p}-1}+\frac{\sin
^{2}(gt)}{e^{q}-1}\right) .  \label{EB}
\end{equation}

\begin{figure}[h]
\begin{center}
\epsfig{figure=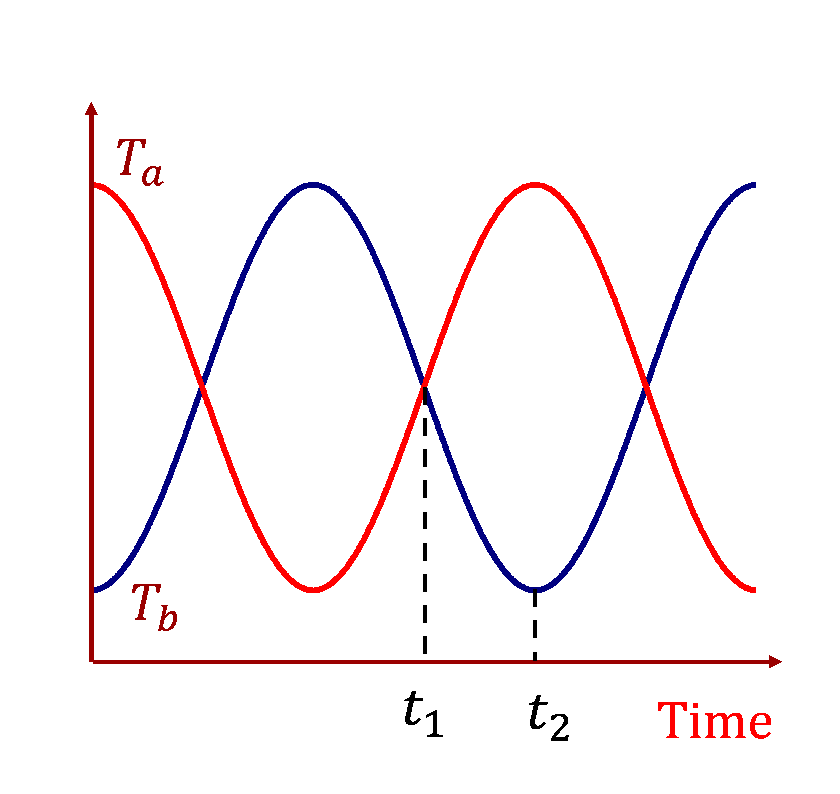, angle=0, width=7cm}
\end{center}
\caption{Effective temperature of oscillators $a$ and $b$ as a function of
time.}
\label{FigQHE-3}
\end{figure}

The energy flow is governed by the correlations induced between oscillators.
At the moments of time when $\cos (gt)=0$ the state of the system becomes
separable, namely, 
\begin{equation}
\hat{\rho}=Ne^{-q\hat{b}^{\dag }\hat{b}}e^{-p\hat{a}^{\dag }\hat{a}}.
\label{p5}
\end{equation}

Comparison of Eq. (\ref{p5}) with the initial state (\ref{q2}) shows that
oscillators swap their thermal states (temperatures), that is hot oscillator
turns cold and vice versa. This feature of quantum evolution appears due to
correlations induced between oscillators in the process of energy exchange.
Please note that in the conventional Rabi oscillation dynamics the
oscillators are usually in Fock states and swap their excitation number
during system's evolution. In the present case the systems exchange the
entire thermal statistics.

Tracing the Gaussian density operator (\ref{p4}) over the oscillator $b$
leaves the oscillator $a$ in a thermal state and vice versa. That is
oscillators can be characterized by an effective temperature at any $t$ even
though the state of the entire system is not thermal. The oscillator
temperatures $T_{a}(t)$ and $T_{b}(t)$ are related to the average oscillator
energies (\ref{EA}) and (\ref{EB}) according to 
\begin{equation}
E_{a}(t)=\frac{\hslash \omega }{e^{\frac{\hslash \omega }{k_{B}T_{a}(t)}}-1}%
,\quad E_{b}(t)=\frac{\hslash \omega }{e^{\frac{\hslash \omega }{%
k_{B}T_{b}(t)}}-1}.
\end{equation}

In Fig. \ref{FigQHE-3} we plot effective temperature of the oscillators $a$
and $b$ as a function of time. During the time interval $t_{1}<t<t_{2}$ the
effective temperature of the colder oscillator decreases while the
temperature of the hotter oscillator increases. The process looks like as if
heat flows from the cold to the hot oscillator.

However, in terms of the collective modes (\ref{M1}) and (\ref{M2}) the
total state is separable at any $t$ (see Eq. (\ref{q3b})) and the average
numbers of excitations $\left\langle \hat{A}^{\dag }(t)\hat{A}%
(t)\right\rangle $ and $\left\langle \hat{B}^{\dag }(t)\hat{B}%
(t)\right\rangle $ are independent of time. In this basis the collective
modes are not coupled and there is no energy exchange between them.

\subsection{Coupled oscillators in coherent state}

Using our approach one can also find analytical solutions for a certain
class of pure states. For example, if initially coupled oscillators are in a
separable coherent state described by the state vector%
\begin{equation*}
\left\vert \psi _{0}\right\rangle =e^{\alpha \hat{a}^{\dag }-\alpha ^{\ast }%
\hat{a}}e^{\beta \hat{b}^{\dag }-\beta ^{\ast }\hat{b}}\left\vert
0_{a}0_{b}\right\rangle ,
\end{equation*}
where $\left\vert 0_{a}0_{b}\right\rangle $ denote the ground state for both
oscillators, the time evolution of the state vector is given by%
\begin{equation*}
\left\vert \psi (t)\right\rangle =e^{\alpha \hat{A}^{\dag }(t)-\alpha ^{\ast
}\hat{A}(t)}e^{\beta \hat{B}^{\dag }(t)-\beta ^{\ast }\hat{B}(t)}\left\vert
0_{a}0_{b}\right\rangle ,
\end{equation*}%
where $\hat{A}(t)$ and $\hat{B}(t)$ are operators obeying the conventional
Heisenberg equation of motion%
\begin{equation*}
i\hslash \frac{\partial \hat{A}(t)}{\partial t}=-\left[ \hat{H},\hat{A}(t)%
\right] ,
\end{equation*}%
\begin{equation*}
i\hslash \frac{\partial \hat{B}(t)}{\partial t}=-\left[ \hat{H},\hat{B}(t)%
\right] ,
\end{equation*}
subject to the initial conditions $\hat{A}(0)=\hat{a}$ and $\hat{B}(0)=\hat{b%
}$.

If oscillators have the same frequency $\omega $ which is independent of
time, and their Hamiltonian is given by Eq. (\ref{q1}), we obtain%
\begin{equation*}
\hat{A}(t)=e^{-i\omega t}\left( \cos (gt)\hat{a}-i\sin (gt)\hat{b}\right) ,
\end{equation*}%
\begin{equation*}
\hat{B}(t)=e^{-i\omega t}\left( \cos (gt)\hat{b}-i\sin (gt)\hat{a}\right) ,
\end{equation*}%
and the time evolution of the system's state vector reads%
\begin{equation*}
\left\vert \psi (t)\right\rangle =e^{\tilde{\alpha}(t)\hat{a}^{\dag }-\tilde{%
\alpha}^{\ast }(t)\hat{a}}e^{\tilde{\beta}(t)\hat{b}^{\dag }-\tilde{\beta}%
^{\ast }(t)\hat{b}}\left\vert 0_{a}0_{b}\right\rangle ,
\end{equation*}%
where%
\begin{equation*}
\tilde{\alpha}(t)=e^{i\omega t}\left( \alpha \cos (gt)+i\beta \sin
(gt)\right) ,
\end{equation*}%
\begin{equation*}
\tilde{\beta}(t)=e^{i\omega t}\left( \beta \cos (gt)+i\alpha \sin
(gt)\right) .
\end{equation*}%
That is oscillators remain in a separable coherent state during evolution
(no entanglement is generated), and the average oscillator excitation
numbers evolve as%
\begin{equation*}
n_{a}(t)=|\tilde{\alpha}(t)|^{2}=|\alpha \cos (gt)+i\beta \sin (gt)|^{2},
\end{equation*}%
\begin{equation*}
n_{b}(t)=|\tilde{\beta}(t)|^{2}=|\beta \cos (gt)+i\alpha \sin (gt)|^{2}.
\end{equation*}%
Writing%
\begin{equation*}
\frac{\beta }{\alpha }=\left\vert \frac{\beta }{\alpha }\right\vert
e^{i\varphi },
\end{equation*}%
where $\varphi $ is the phase difference between $\beta $ and $\alpha $, we
find%
\begin{equation*}
n_{a}(t)=n_{a}(0)\cos ^{2}(gt)+n_{b}(0)\sin ^{2}(gt)
\end{equation*}%
\begin{equation}
-\sin \varphi \sqrt{n_{a}(0)n_{b}(0)}\sin (2gt),  \label{QA1}
\end{equation}%
\begin{equation*}
n_{b}(t)=n_{b}(0)\cos ^{2}(gt)+n_{a}(0)\sin ^{2}(gt)
\end{equation*}%
\begin{equation}
+\sin \varphi \sqrt{n_{a}(0)n_{b}(0)}\sin (2gt).  \label{QA2}
\end{equation}

\begin{figure}[h]
\begin{center}
\epsfig{figure=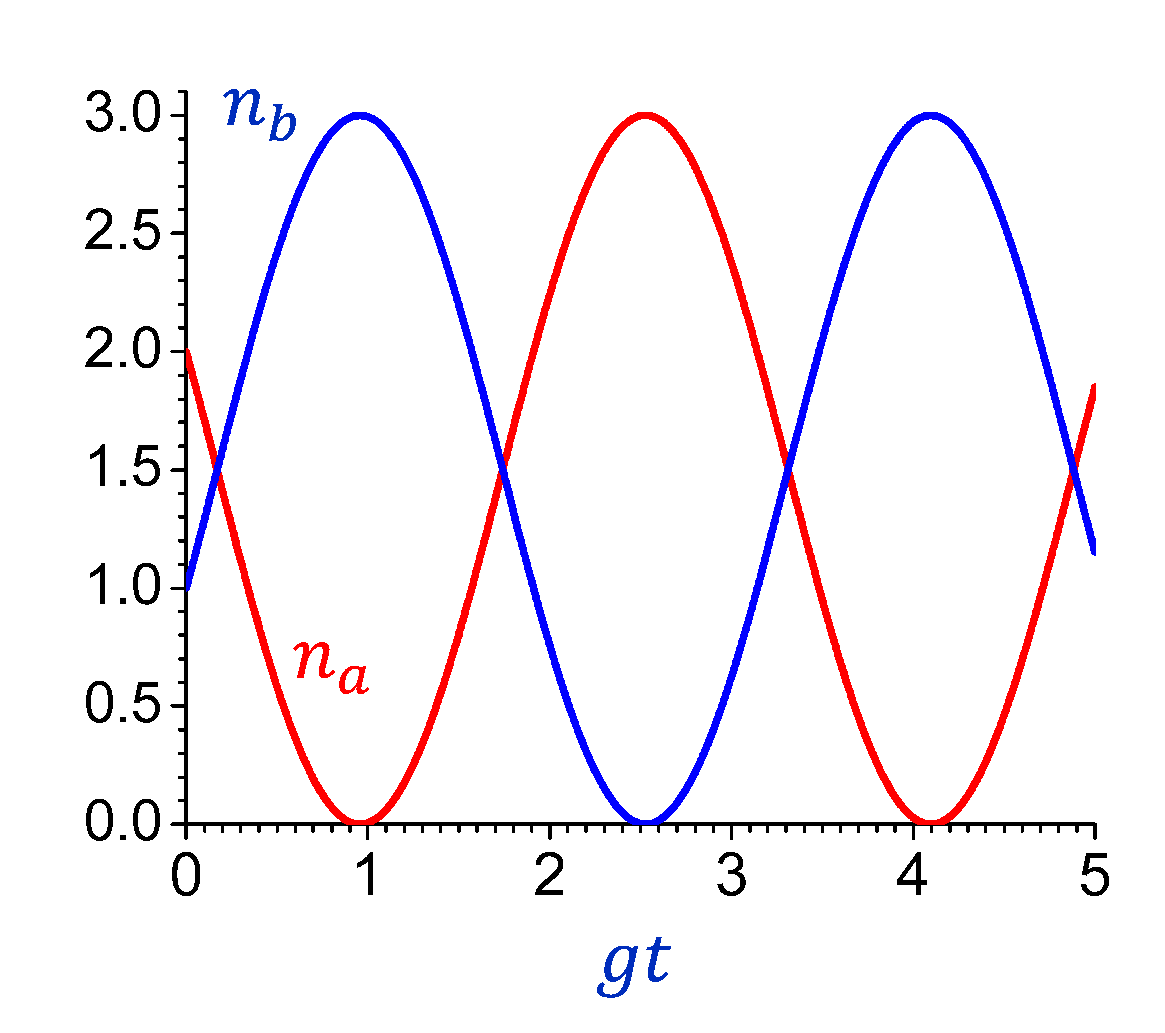, angle=0, width=7cm}
\end{center}
\caption{Average number of oscillator excitations as a function of time.
Initially oscillators are prepared in a coherent state with relative phase $%
\protect\varphi =\protect\pi /2$. During evolution all population of the
oscillator $a$ is transferred to $b$ and vice versa. }
\label{Fig2b}
\end{figure}

In particular, if $\varphi =0$ or $\varphi =\pi $ the time evolution for $%
n_{a}(t)$ and $n_{b}(t)$ is the same as in the case of initially thermal
states and oscillators swap their excitation numbers. However, if $\varphi
=\pi /2$ then%
\begin{equation*}
n_{a}(t)=\left( \sqrt{n_{a}(0)}\cos (gt)-\sqrt{n_{b}(0)}\sin (gt)\right)
^{2},
\end{equation*}%
\begin{equation*}
n_{b}(t)=\left( \sqrt{n_{b}(0)}\cos (gt)+\sqrt{n_{a}(0)}\sin (gt)\right)
^{2},
\end{equation*}%
and all population of the oscillator $a$ is transferred to $b$ and vice
versa (see Fig. \ref{Fig2b}).

Quantum evolution of coupled oscillators in coherent states is analogous to
evolution of classical coupled oscillators $A$ and $B$ described by equations%
\begin{equation*}
\ddot{A}+\omega ^{2}A=2\omega gB,\quad \ddot{B}+\omega ^{2}B=2\omega gA,
\end{equation*}%
which for $g\ll \omega $ have approximate solution%
\begin{equation*}
A(t)=e^{-i\omega t}\left[ \cos (gt)A(0)-i\sin (gt)B(0)\right] ,
\end{equation*}%
\begin{equation*}
B(t)=e^{-i\omega t}\left[ \cos (gt)B(0)-i\sin (gt)A(0)\right] .
\end{equation*}%
Energies of such classical oscillators are proportional to $n_{a}(t)$ and $%
n_{b}(t)$ given by Eqs. (\ref{QA1}) and (\ref{QA2}). In the present case, $%
\varphi $ is the difference between oscillator phases at the initial moment
of time $B(0)/A(0)=\left\vert B(0)/A(0)\right\vert e^{i\varphi }$.

One should note that present formalism allows us to find quantum evolution
for any initial Gaussian state (see Sect. \ref{Gaussian}). In particular, a
two-mode squeezed state%
\begin{equation}
\left\vert \psi _{0}\right\rangle =e^{\gamma \hat{a}\hat{b}-\gamma ^{\ast }%
\hat{a}^{\dag }\hat{b}^{\dag }}\left\vert 0_{a}0_{b}\right\rangle ,
\label{QA3}
\end{equation}
under Hamiltonian (\ref{q1}) with $\omega _{a}=\omega _{b}=\omega $, evolves
as%
\begin{equation*}
\left\vert \psi (t)\right\rangle =e^{\gamma e^{-2i\omega t}\left[ \cos (2gt)%
\hat{a}\hat{b}-\frac{i}{2}\sin (2gt)\left( \hat{a}^{2}+\hat{b}^{2}\right) %
\right] -h.c.}\left\vert 0_{a}0_{b}\right\rangle ,
\end{equation*}%
and at the time moments for which $\sin (2gt)=\pm 1$ it becomes a separable
product of single-mode squeezed states%
\begin{equation}
\left\vert \psi \right\rangle =e^{\mp \frac{i\gamma }{2}e^{-2i\omega t}\hat{a%
}^{2}\pm h.c.}e^{\mp \frac{i\gamma }{2}e^{-2i\omega t}\hat{b}^{2}\pm
h.c.}\left\vert 0_{a}0_{b}\right\rangle .  \label{QA4}
\end{equation}

Tracing over oscillator $\hat{b}$ leaves the initial two-mode squeezed state
(\ref{QA3}) in a thermal state for the oscillator $\hat{a}$, and vice versa.
That is initial statistics of each oscillator is thermal. Equation (\ref{QA4}%
) shows that during system's evolution \textquotedblleft
thermal\textquotedblright\ oscillators become squeezed due to initial
correlations between oscillators. In quantum optics, states (\ref{QA4}) are
known as two-photon coherent states \cite{Yuen76}.

\subsection{Quantum evolution of coupled two-level atoms}

Next we consider another system for which time evolution of the density
operator can be found analytically. Namely, we consider two coupled
two-level atoms $a$ and $b$ with equal transition frequencies $\omega $
described by the Hamiltonian 
\begin{equation}
\hat{H}=\frac{1}{2}\hslash \omega \hat{\sigma}_{z}+\frac{1}{2}\hslash \omega 
\hat{\pi}_{z}+\hslash g\left( \hat{\sigma}^{\dag }\hat{\pi}+\hat{\sigma}\hat{%
\pi}^{\dag }\right) ,
\end{equation}%
where atomic operators obey commutation relations%
\begin{equation}
\lbrack \hat{\sigma},\hat{\sigma}^{\dag }]=-\hat{\sigma}_{z},\quad \lbrack 
\hat{\sigma},\hat{\sigma}_{z}]=2\hat{\sigma},\quad \lbrack \hat{\sigma}%
^{\dag },\hat{\sigma}_{z}]=-2\hat{\sigma}^{\dag },  \label{CR1}
\end{equation}%
and similar relations hold for $\hat{\pi}$. If initially the system's
density operator is thermal%
\begin{equation}
\hat{\rho}_{0}(\hat{\sigma}_{z},\hat{\pi}_{z})=Ne^{-q\hat{\sigma}_{z}/2}e^{-p%
\hat{\pi}_{z}/2},
\end{equation}%
where $q=\hslash \omega /k_{B}T_{a}$, $p=\hslash \omega /k_{B}T_{b}$, 
\begin{equation*}
N=\frac{4}{\cosh \left( q/2\right) \cosh (p/2)},
\end{equation*}%
it evolves as 
\begin{equation}
\hat{\rho}(t)=Ne^{-q\hat{\Xi}_{z}(t)/2}e^{-p\hat{\Pi}_{z}(t)/2},
\end{equation}%
where operators $\hat{\Xi}_{z}(t)$ and $\hat{\Pi}_{z}(t)$ obey evolution
equations%
\begin{equation}
i\hslash \frac{\partial \hat{\Xi}_{z}(t)}{\partial t}=\left[ \hat{H},\hat{\Xi%
}_{z}(t)\right] ,  \label{QQQ1}
\end{equation}%
\begin{equation}
i\hslash \frac{\partial \hat{\Pi}_{z}(t)}{\partial t}=\left[ \hat{H},\hat{\Pi%
}_{z}(t)\right]  \label{QQQ2}
\end{equation}%
subject to the initial conditions $\hat{\Xi}_{z}(0)=\hat{\sigma}_{z}$ and $%
\hat{\Pi}_{z}(0)=\hat{\pi}_{z}$. Evolution Eqs. (\ref{QQQ1}) and (\ref{QQQ2}%
) yield%
\begin{equation}
\hat{\Xi}_{z}(t)+\hat{\Pi}_{z}(t)=\hat{\sigma}_{z}+\hat{\pi}_{z}.
\label{CCC1}
\end{equation}%
Taking into account that for two-level systems 
\begin{equation*}
\hat{\Xi}(t)\hat{\Xi}^{\dag }(t)+\hat{\Xi}^{\dag }(t)\hat{\Xi}(t)=1,
\end{equation*}%
\begin{equation*}
\hat{\Pi}(t)\hat{\Pi}^{\dag }(t)+\hat{\Pi}^{\dag }(t)\hat{\Pi}(t)=1,
\end{equation*}%
we obtain 
\begin{equation*}
\frac{d^{2}}{dt^{2}}\left( \hat{\Xi}_{z}(t)-\hat{\Pi}_{z}(t)\right)
+4g^{2}\left( \hat{\Xi}_{z}(t)-\hat{\Pi}_{z}(t)\right) =0.
\end{equation*}%
Combining this with Eq. (\ref{CCC1}) we find%
\begin{equation*}
\hat{\Xi}_{z}(t)=\frac{1}{2}\left( \hat{\sigma}_{z}+\hat{\pi}_{z}\right) +%
\frac{1}{2}\left( \hat{\sigma}_{z}-\hat{\pi}_{z}\right) \cos (2gt)
\end{equation*}%
\begin{equation*}
+i\left( \hat{\pi}\hat{\sigma}^{\dag }-\hat{\pi}^{\dag }\hat{\sigma}\right)
\sin (2gt),
\end{equation*}

\begin{equation*}
\hat{\Pi}_{z}(t)=\frac{1}{2}\left( \hat{\sigma}_{z}+\hat{\pi}_{z}\right) -%
\frac{1}{2}\left( \hat{\sigma}_{z}-\hat{\pi}_{z}\right) \cos (2gt)
\end{equation*}%
\begin{equation*}
-i\left( \hat{\pi}\hat{\sigma}^{\dag }-\hat{\pi}^{\dag }\hat{\sigma}\right)
\sin (2gt).
\end{equation*}%
That is system undergoes Rabi oscillations with frequency $2g$. For the
average numbers of excitations of the atoms $n_{a}(t)$ and $n_{b}(t)$ we
obtain the same expression as for two harmonic oscillators initially
prepared in the thermal state.

\section{Noise induced coherence}

\label{Sec3}

Next, we consider three oscillators $\hat{a}$, $\hat{b}$ and $\hat{c}$
having the same frequency $\omega $. Oscillators $\hat{b}$ and $\hat{c}$ are
coupled with $\hat{a}$, but not with each other. The Hamiltonian of the
system reads%
\begin{equation}
\hat{H}=\hslash \omega \left( \hat{a}^{\dag }\hat{a}+\hat{b}^{\dag }\hat{b}+%
\hat{c}^{\dag }\hat{c}\right) +\hslash g\left( \hat{a}^{\dag }(\hat{b}+\hat{c%
})+\hat{a}(\hat{b}^{\dag }+\hat{c}^{\dag })\right) .  \label{s12a}
\end{equation}%
We assume that initially the oscillator $\hat{a}$ is in a thermal state with
hot temperature $T_{h}$, while oscillators $\hat{b}$ and $\hat{c}$ have cold
temperature $T_{c}$. The initial density operator of the system is%
\begin{equation}
\hat{\rho}_{0}(\hat{a},\hat{b},\hat{c})=Ne^{-q\hat{a}^{\dag }\hat{a}%
}e^{-p\left( \hat{b}^{\dag }\hat{b}+\hat{c}^{\dag }\hat{c}\right) },
\label{s12b}
\end{equation}%
where $q=\hslash \omega /k_{B}T_{h}$, $p=\hslash \omega /k_{B}T_{c}$ and $N$
is the normalization factor%
\begin{equation*}
N=\left( 1-e^{-q}\right) \left( 1-e^{-p}\right) ^{2}.
\end{equation*}%
In the present problem the oscillator $\hat{a}$ serves as an incoherent
energy source.

Introducing collective operators 
\begin{equation}
\hat{B}_{1}=\frac{1}{\sqrt{2}}\left( \hat{b}+\hat{c}\right) ,\quad \hat{B}%
_{2}=\frac{1}{\sqrt{2}}\left( \hat{b}-\hat{c}\right) ,
\end{equation}%
which obey bosonic commutation relations, one can write the system's
Hamiltonian as%
\begin{equation}
\hat{H}=\hslash \omega \left( \hat{a}^{\dag }\hat{a}+\hat{B}_{1}^{\dag }\hat{%
B}_{1}+\hat{B}_{2}^{\dag }\hat{B}_{2}\right) +\sqrt{2}\hslash g\left( \hat{a}%
^{\dag }\hat{B}_{1}+\hat{a}\hat{B}_{1}^{\dag }\right) ,  \label{n1}
\end{equation}%
while%
\begin{equation}
\hat{\rho}_{0}(\hat{a},\hat{B}_{1},\hat{B}_{2})=Ne^{-q\hat{a}^{\dag }\hat{a}%
}e^{-p\left( \hat{B}_{1}^{\dag }\hat{B}_{1}+\hat{B}_{2}^{\dag }\hat{B}%
_{2}\right) }.  \label{n2}
\end{equation}

Equation (\ref{n1}) shows that only symmetric collective mode\ $\hat{B}_{1}$
is coupled with $\hat{a}$ and the effective coupling constant is $\sqrt{2}g$%
. Thus, the problem reduces to two coupled oscillators ($\hat{a}$ and $\hat{B%
}_{1}$) of the previous section, while the antisymmetric mode\ $\hat{B}_{2}$
remains in the thermal state with temperature $T_{c}$ during the system's
evolution. Using the previous result, we obtain the time evolution of the
system's density operator%
\begin{equation*}
\hat{\rho}(t)=Ne^{-p\hat{B}_{2}^{\dag }\hat{B}_{2}}
\end{equation*}
\begin{equation*}
\times e^{-q\left( \cos (\sqrt{2}gt)\hat{a}^{\dag }-i\sin (\sqrt{2}gt)\hat{B}%
_{1}^{\dag }\right) \left( \cos (\sqrt{2}gt)\hat{a}+i\sin (\sqrt{2}gt)\hat{B}%
_{1}\right) }
\end{equation*}
\begin{equation*}
\times e^{-p\left( \cos (\sqrt{2}gt)\hat{B}_{1}^{\dag }-i\sin (\sqrt{2}gt)%
\hat{a}^{\dag }\right) \left( \cos (\sqrt{2}gt)\hat{B}_{1}+i\sin (\sqrt{2}gt)%
\hat{a}\right) }.
\end{equation*}%
At the moments of time when $\cos (\sqrt{2}gt)=0$ the state of the system
becomes 
\begin{equation}
\hat{\rho}=Ne^{-q\hat{B}_{1}^{\dag }\hat{B}_{1}}e^{-p\hat{a}^{\dag }\hat{a}%
}e^{-p\hat{B}_{2}^{\dag }\hat{B}_{2}},  \label{n4}
\end{equation}%
that is oscillators $\hat{a}$ and $\hat{B}_{1}$ swap temperatures. In terms
of the original operators $\hat{b}$ and $\hat{c}$, Eq. (\ref{n4}) reads%
\begin{equation}
\hat{\rho}=Ne^{\frac{p-q}{2}(\hat{b}^{\dag }+\hat{c}^{\dag })(\hat{b}+\hat{c}%
)}e^{-p\left( \hat{b}^{\dag }\hat{b}+\hat{c}^{\dag }\hat{c}\right) }e^{-p%
\hat{a}^{\dag }\hat{a}}.  \label{n5}
\end{equation}%
Equation (\ref{n5}) shows that if $T_{h}\neq T_{c}$ (that is $q\neq p$)
oscillators $\hat{b}$ and $\hat{c}$ become correlated. In particular, using
Eq. (\ref{n4}), we obtain%
\begin{equation*}
\left\langle b\hat{c}^{\dag }\right\rangle =\frac{1}{2}\text{Tr}\left( (\hat{%
B}_{1}+\hat{B}_{2})(\hat{B}_{1}^{\dag }-\hat{B}_{2}^{\dag })\hat{\rho}\right)
\end{equation*}%
\begin{equation*}
=\frac{1}{2}\left( \frac{1}{e^{q}-1}-\frac{1}{e^{p}-1}\right) .
\end{equation*}%
That is to say the interaction with the same incoherent energy source $\hat{a%
}$ induces coherence between oscillators $\hat{b}$ and $\hat{c}$. Such noise
induced coherence can be used to enhance performance of solar cells \cite%
{Svid11,Svid12} and could play a role in biological processes \cite%
{Enge07,Lee07,Calh09}.

Thermal noise is implicitly present in our problem. It is needed to prepare
the initial state of the system as a thermal state with two temperatures. To
prepare such initial state oscillators must be connected with thermal baths.
The latter are noisy sources. During subsequent evolution the system is
isolated but coherence emerges out of the initial \textquotedblleft
noisy\textquotedblright\ non-equilibrium state.

\begin{figure}[h]
\begin{center}
\epsfig{figure=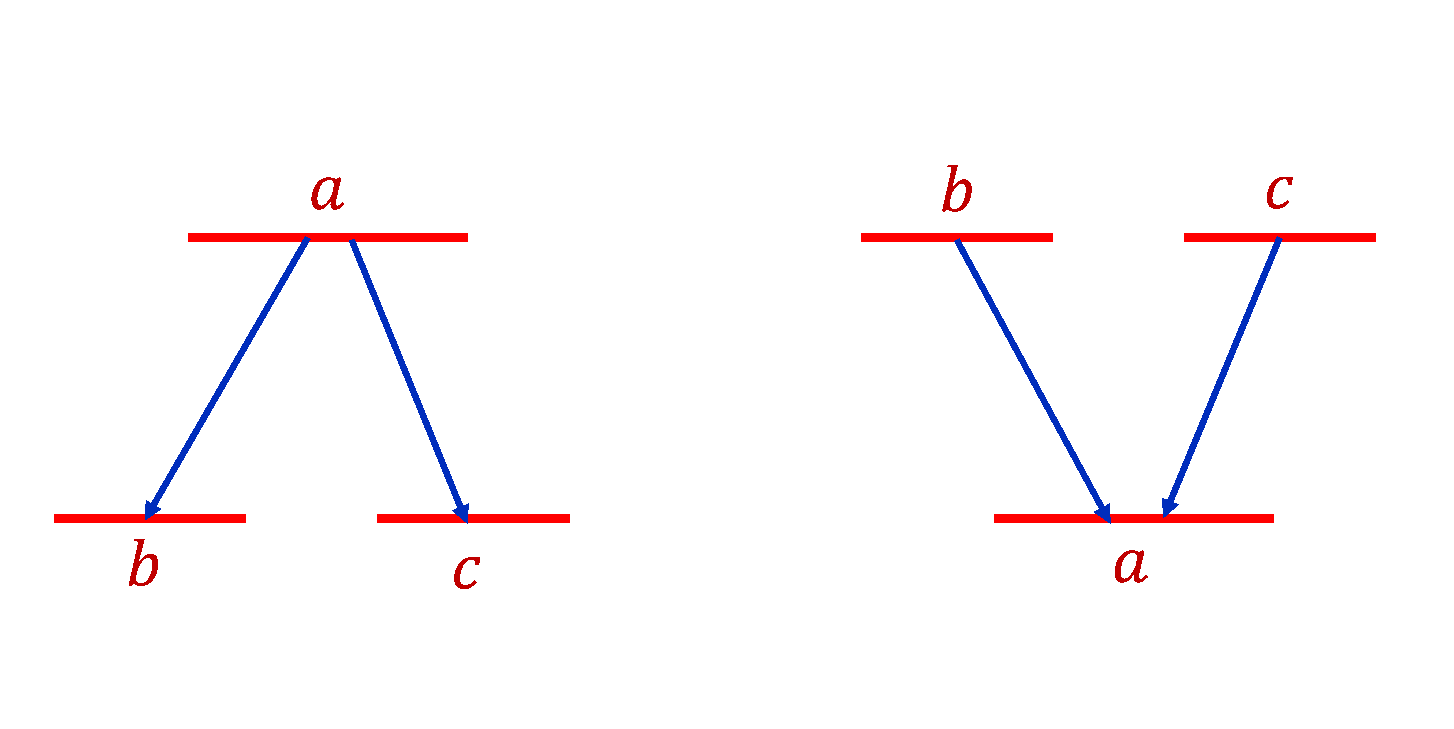, angle=0, width=8cm}
\end{center}
\caption{ Radiative decay from (into) a common level into (from) degenerate
levels induce coherence between degenerate levels.}
\label{Fig1aa}
\end{figure}

The latter process is somewhat analogous to the Fano-induced coherence when
a three-level atom spontaneously decays into a degenerate ground state or
into a common ground state level from degenerate excited states as shown in
Fig. \ref{Fig1aa}. This process generates coherence between degenerate
levels \cite{Svid11}. The scheme sketched in the left part of Fig. \ref%
{Fig1aa} is analogous to our three oscillator model in which oscillator $%
\hat{a}$ is hotter than oscillators $\hat{b}$ and $\hat{c}$. While the right
scheme is analogous to the case when oscillator $\hat{a}$ is cooler than
oscillators $\hat{b}$ and $\hat{c}$, and thus energy flows to the oscillator 
$\hat{a}$. In the latter case the oscillator $\hat{a}$ serves as an
incoherent energy sink.

Equation (\ref{n4}) shows that the correlated state of oscillators $\hat{b}$
and $\hat{c}$ is separable in terms of the collective modes $\hat{B}_{1}$
and $\hat{B}_{2}$, namely, it is a product of thermal states with different
temperatures $T_{h}$ and $T_{c}$%
\begin{equation*}
e^{\frac{p-q}{2}(\hat{b}^{\dag }+\hat{c}^{\dag })(\hat{b}+\hat{c}%
)}e^{-p\left( \hat{b}^{\dag }\hat{b}+\hat{c}^{\dag }\hat{c}\right) }=e^{-q%
\hat{B}_{1}^{\dag }\hat{B}_{1}}e^{-p\hat{B}_{2}^{\dag }\hat{B}_{2}}.
\end{equation*}%
Thus, one can extract work from the state with quantum coherence (\ref{n5})
in just the same way as if QHE operates between pair of hot and cold thermal
reservoirs with temperatures $T_{h}$ and $T_{c}$. To prepare state (\ref{n5}%
) extra work is required and, hence, there is no violation of the Second law
of thermodynamics.

\section{Photonic quantum heat engine}

\label{Sec4}

\begin{figure}[h]
\begin{center}
\epsfig{figure=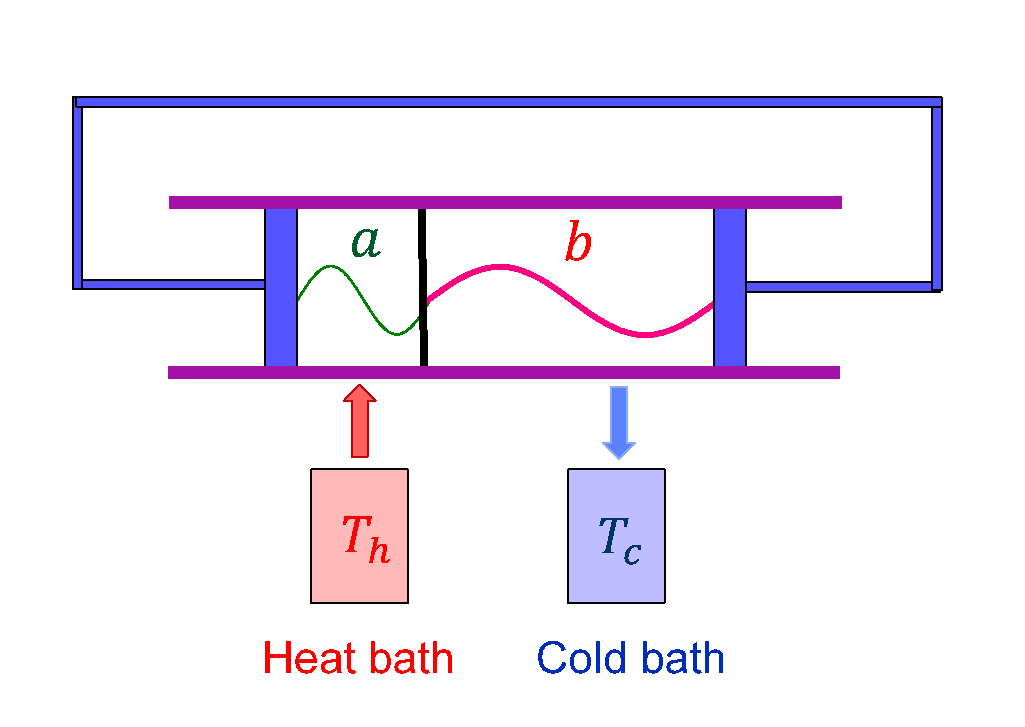, angle=0, width=9cm}
\end{center}
\caption{Photonic quantum heat engine composed of two optical cavities with
different frequencies coupled via a partially transmitting common mirror.
Outside mirrors can move and act as a piston of the engine. The working
fluid of the engine is photon gas in the cavities which exerts radiation
pressure on the mirrors and can do mechanical work on the surroundings. }
\label{Fig1}
\end{figure}

In this section, we consider a quantum heat engine in which two oscillators
described by the Hamiltonian (\ref{q1}) with different frequencies $\omega
_{a}$ and $\omega _{b}$ ($\omega _{b}<\omega _{a}$) serve as the working
substance. Photon gas in optical cavities of different lengths can be
modeled as two coupled harmonic oscillators (see Fig. \ref{Fig1}). In our
setup, photons can tunnel between the cavities through a partially
transmitting common mirror which yields coupling between the oscillators.
Outside mirrors can move and act as a piston of the engine. Photon gas
exerts radiation pressure on the mirrors and can do mechanical work on the
surroundings. Frequencies of photons in the cavities depend on the cavity
size and time due to varying positions of the outside mirrors.

At the beginning of the engine cycle, the cavities are not coupled. The
cavity $a$ is in contact with a heat reservoir with temperature $T_{h}$,
while cavity $b$ is in contact with a cold reservoir with temperature $T_{c}$%
. Thus, initially, photons in the cavities are in the thermal state (\ref{q2}%
) with $q=\hslash \omega _{a}/k_{B}T_{h}$ and $p=\hslash \omega
_{b}/k_{B}T_{c}$. Then, cavities are disconnected from the reservoirs and
become coupled with each other through the partially transmitting common
mirror. At this part of the cycle, the working substance is an isolated
system; the piston is moving and the oscillator (photon) frequencies vary
with time forcing the working substance to do mechanical work $W$ on the
surroundings.

For the moment, we will not specify how the piston moves and consider a
general situation. The density operator of the coupled oscillators (photons
in the two cavities) evolves according to Eqs. (\ref{q3b}), (\ref{y0})-(\ref%
{y2a}). At the end of the cycle, the cavity mirrors return to their initial
positions and the oscillator frequencies return to their initial values $%
\omega _{a}$ and $\omega _{b}$.

To find the efficiency of the engine, we note that operator $\hat{a}^{\dag }%
\hat{a}+\hat{b}^{\dag }\hat{b}$ commutes with the Hamiltonian (\ref{q1}) and
thus, the total number of the oscillator excitations remains constant%
\begin{equation}
n_{a}(t)+n_{b}(t)=n_{a}(0)+n_{b}(0),  \label{c1}
\end{equation}%
where%
\begin{equation}
n_{a}(t)=\text{Tr}\left( \hat{a}^{\dag }\hat{a}\hat{\rho}(t)\right) ,\quad
n_{b}(t)=\text{Tr}(\hat{b}^{\dag }\hat{b}\hat{\rho}(t)).  \label{v1}
\end{equation}%
Work done by the engine is equal to the difference in energy of the
oscillators at the beginning and the end of the cycle. Since at the end of
the cycle (at $t=t_{c}$) the oscillator frequencies return to their initial
values $\omega _{a}$ and $\omega _{b}$, we obtain that the work done by the
system during one cycle is%
\begin{equation*}
W=\hslash \omega _{a}\left( n_{a}(0)-n_{a}(t_{c})\right) +\hslash \omega
_{b}\left( n_{b}(0)-n_{b}(t_{c})\right) .
\end{equation*}%
Using Eq. (\ref{c1}), we find%
\begin{equation}
W=\hslash \left( \omega _{a}-\omega _{b}\right) \left(
n_{a}(0)-n_{a}(t_{c})\right) .  \label{c1a}
\end{equation}

At the end of the cycle, the oscillators become decoupled and are brought
into contact with the heat and cold baths. Heat taken from the hot reservoir
is%
\begin{equation*}
Q=\hslash \omega _{a}\left( n_{a}(0)-n_{a}(t_{c})\right) ,
\end{equation*}%
and hence, the engine efficiency is given by%
\begin{equation}
\eta =\frac{W}{Q}=1-\frac{\omega _{b}}{\omega _{a}}.  \label{c2}
\end{equation}%
Equation (\ref{c2}) shows that the engine efficiency depends only on the
initial frequencies $\omega _{a}$ and $\omega _{b}$ and is independent of
how oscillator frequencies vary (piston moves) during the engine operation
(see also Ref. \cite{Scov59}). However, the work done by the engine is
positive not for all values of $\omega _{a}$ and $\omega _{b}$. To find the
sign of $W,$ it is convenient to express operator $\hat{a}$ in terms of $%
\hat{A}(t)$ and $\hat{B}(t)$ using Eqs. (\ref{y0}) which yields 
\begin{equation}
\hat{a}=c(t)\hat{A}(t)+d(t)\hat{B}(t).  \label{p6}
\end{equation}%
Since operators $\hat{a}$ and $\hat{a}^{\dag }$ obey the same bosonic
commutation relations as $\hat{A}(t)$ and $\hat{A}^{\dag }(t)$, the
time-dependent functions in Eq. (\ref{p6}) have the following integral of
motion 
\begin{equation}
|c(t)|^{2}+|d(t)|^{2}=1.  \label{p8}
\end{equation}

Using Eq. (\ref{p6}) and the formula for the density operator (\ref{q3b}),
we find%
\begin{equation*}
n_{a}(t)=\text{Tr}\left[ \hat{a}^{\dag }\hat{a}\hat{\rho}(t)\right] =
\end{equation*}%
\begin{equation*}
\text{Tr}\left[ \left( |c(t)|^{2}\hat{A}^{\dag }(t)\hat{A}(t)+|d(t)|^{2}\hat{%
B}^{\dag }(t)\hat{B}(t)\right) \hat{\rho}(t)\right] =
\end{equation*}%
\begin{equation*}
=\frac{|c(t)|^{2}}{e^{q}-1}+\frac{|d(t)|^{2}}{e^{p}-1}.
\end{equation*}%
Taking into account Eqs. (\ref{c1a}) and (\ref{p8}), we obtain

\begin{equation}
W=\hslash \left( \omega _{a}-\omega _{b}\right) \left( \frac{1}{e^{q}-1}-%
\frac{1}{e^{p}-1}\right) |d(t_{c})|^{2}.  \label{p11}
\end{equation}

According to Eq. (\ref{p11}), the work done by the engine is positive
provided $q<p$, that is 
\begin{equation}
\omega _{b}>\frac{T_{c}}{T_{h}}\omega _{a}.  \label{p13}
\end{equation}%
Equations (\ref{c2}) and (\ref{p13}) yield that maximum efficiency of the
QHE is given by the Carnot formula%
\begin{equation}
\eta _{\max }=1-\frac{T_{c}}{T_{h}},  \label{s5}
\end{equation}%
which is achieved for $\omega _{b}=\omega _{a}T_{c}/T_{h}$. According to Eq.
(\ref{p11}), at maximum efficiency the work done by the engine vanishes, $%
W\rightarrow 0$, but the ratio $W/Q$ remains finite. This is consistent with
the operation of classical heat engines at Carnot efficiency which is
achieved when the engine follows a reversible cycle that requires the engine
to run infinitely slowly. As a consequence, the engine's power becomes close
to zero. In contrast, the present quantum cycle has a finite duration and
the QHE's power can vanish only if the work done per cycle goes to zero.

Next, we find how the piston should move to produce maximum work per cycle
of the engine operation at given initial frequencies $\omega _{a}$ and $%
\omega _{b}$. According to Eq. (\ref{p11}), the work is maximum if $%
|d(t_{c})|=1$. In other words, at the end of the cycle, upto a phase factor, 
$\hat{B}(t_{c})=\hat{a}$ and $\hat{A}(t_{c})=\hat{b}$. Hence%
\begin{equation}
\hat{\rho}(t_{c})=Ne^{-q\hat{b}^{\dag }\hat{b}}e^{-p\hat{a}^{\dag }\hat{a}},
\end{equation}%
and at the end of the cycle the oscillators swap numbers of excitations. At
this moment, the oscillator temperatures are%
\begin{equation*}
T_{a}=\frac{\omega _{a}}{\omega _{b}}T_{c},\quad T_{b}=\frac{\omega _{b}}{%
\omega _{a}}T_{h}.
\end{equation*}

Thus, we need to find solution of Eqs. (\ref{y1})-(\ref{y2a}) satisfying
condition%
\begin{equation}
C(t_{c})=0,\quad |D(t_{c})|=1,\quad E(t_{c})=0,\quad |F(t_{c})|=1.
\label{y2b}
\end{equation}%
According to Eqs. (\ref{p1}), (\ref{p2}) and (\ref{p3}), the last three
conditions in Eq. (\ref{y2b}) follow from the first one. Hence, we only need
to find $C(t)$. Introducing new functions $\tilde{C}$ and $\tilde{D}$ such
that 
\begin{equation*}
C=e^{\frac{i}{2}\int_{0}^{t}dt^{\prime }(\omega _{a}(t^{\prime })+\omega
_{b}(t))}\tilde{C},\quad D=e^{\frac{i}{2}\int_{0}^{t}dt^{\prime }(\omega
_{a}(t^{\prime })+\omega _{b}(t))}\tilde{D},
\end{equation*}%
Eqs. (\ref{y1}) reduce to%
\begin{equation}
\frac{d^{2}\tilde{C}}{dt^{2}}+\left( g^{2}+\frac{\Delta ^{2}}{4}-\frac{i}{2}%
\frac{d\Delta }{dt}\right) \tilde{C}=0,  \label{s12}
\end{equation}%
where%
\begin{equation*}
\Delta (t)=\omega _{a}(t)-\omega _{b}(t)
\end{equation*}%
and $\tilde{C}(t)$ satisfies conditions 
\begin{equation}
\tilde{C}(0)=1,\quad \frac{d\tilde{C}(0)}{dt}=\frac{i}{2}\left( \omega
_{a}-\omega _{b}\right) ,\quad \tilde{C}(t_{c})=0.
\end{equation}%
Assuming that%
\begin{equation}
\Delta (t)=\omega _{a}-\omega _{b}+2\delta \sin (\nu t),  \label{s13}
\end{equation}%
where the modulation amplitude $\delta \ll 1$, and disregarding terms of the
order of $\delta ^{2}$, we obtain Mathieu's differential equation for $%
\tilde{C}$ 
\begin{equation*}
\frac{d^{2}\tilde{C}}{dt^{2}}+\left[ g^{2}+\frac{(\omega _{a}-\omega
_{b})^{2}}{4}+\right.
\end{equation*}%
\begin{equation}
\left. \delta \left( \left( \omega _{a}-\omega _{b}\right) \sin (\nu t)-i\nu
\cos (\nu t)\right) \right] \tilde{C}=0.  \label{k2}
\end{equation}%
At parametric resonance, $\nu =\sqrt{4g^{2}+(\omega _{a}-\omega _{b})^{2}}$,
Eq. (\ref{k2}) has the following approximate solution 
\begin{equation*}
\tilde{C}(t)\approx C_{1}(t)e^{i\nu t/2}+C_{2}(t)e^{-i\nu t/2},
\end{equation*}%
where%
\begin{equation}
C_{1}(t)=\frac{1}{2}\left( 1+\frac{\Delta _{0}}{\nu }\right) \cos \left( 
\frac{\delta g}{\nu }t\right) +\frac{g}{\nu }\sin \left( \frac{\delta g}{\nu 
}t\right) ,
\end{equation}%
\begin{equation}
C_{2}(t)=\frac{1}{2}\left( 1-\frac{\Delta _{0}}{\nu }\right) \cos \left( 
\frac{\delta g}{\nu }t\right) -\frac{g}{\nu }\sin \left( \frac{\delta g}{\nu 
}t\right) ,
\end{equation}%
are slowly varying functions of time and $\Delta _{0}=\omega _{a}-\omega
_{b} $. Condition $\tilde{C}(t_{c})=0$ is approximately satisfied if we
choose the cycle duration $t_{c}$ such that $\cos \left( \frac{\delta g}{\nu 
}t_{c}\right) \approx 0$, that is $t_{c}\approx \pi \nu /2\delta g$. Then,
near the end of the cycle 
\begin{equation}
|C(t)|^{2}\approx \frac{4g^{2}}{\nu ^{2}}\cos ^{2}\left( \frac{\nu t}{2}%
\right) ,
\end{equation}%
which vanishes at the minimum of the fast oscillating carrier function.

\begin{figure}[h]
\begin{center}
\epsfig{figure=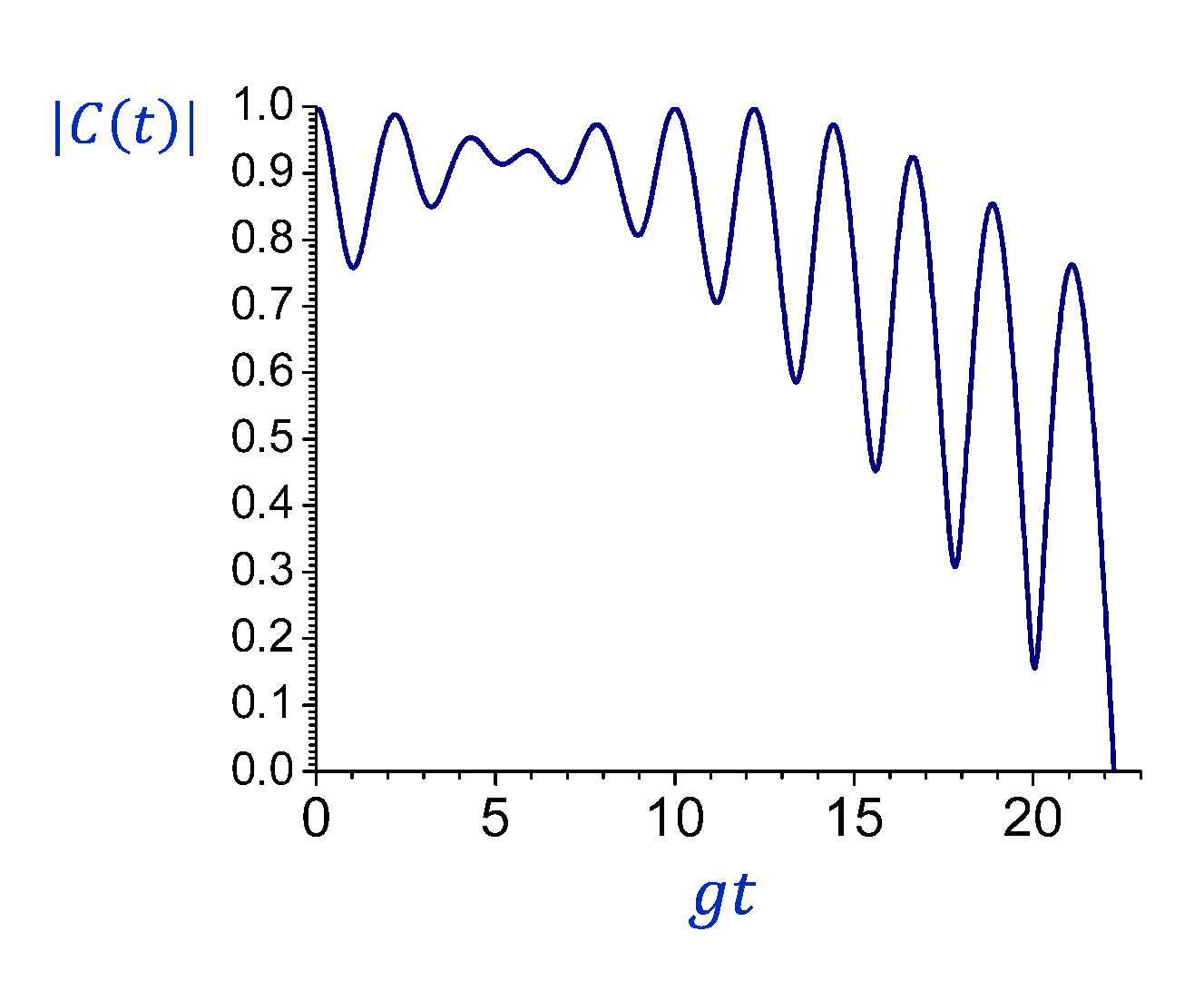, angle=0, width=9cm}
\end{center}
\caption{$|C(t)|$ for one cycle of engine operation obtained by numerical
solution of Eq. (\protect\ref{s12}) with $\protect\omega _{a}-\protect\omega %
_{b}=2g$, $\protect\delta =0.2g$ and $\protect\nu =\protect\sqrt{4g^{2}+(%
\protect\omega _{a}-\protect\omega _{b})^{2}}$. }
\label{Fig2}
\end{figure}

In Fig. (\ref{Fig2}), we plot $|C(t)|$ for one cycle of engine operation
obtained by numerical solution of Eq. (\ref{s12}) in which $\Delta (t)$ is
given by Eq. (\ref{s13}) with $\omega _{a}-\omega _{b}=2g$, $\delta =0.2g$
and the modulation frequency $\nu $ satisfies the condition of parametric
resonance $\nu =\sqrt{4g^{2}+(\omega _{a}-\omega _{b})^{2}}$. For these
parameters, $|C(t)|$ becomes equal to zero at $t_{c}=22.26g$ and the
oscillators swap numbers of excitations at the end of the cycle.

Next we discuss at which time the work protocol has to be stopped to
maximize the engine's power. Straightforward calculations yield that work
per engine cycle is given by

\begin{equation}
W=\hbar (\omega _{a}-\omega _{b})(1-|C(t)|^{2})\left( \frac{1}{e^{q}-1}-%
\frac{1}{e^{p}-1}\right) .
\end{equation}

\begin{figure}[htbp]
\centering
\includegraphics[width=1\columnwidth]{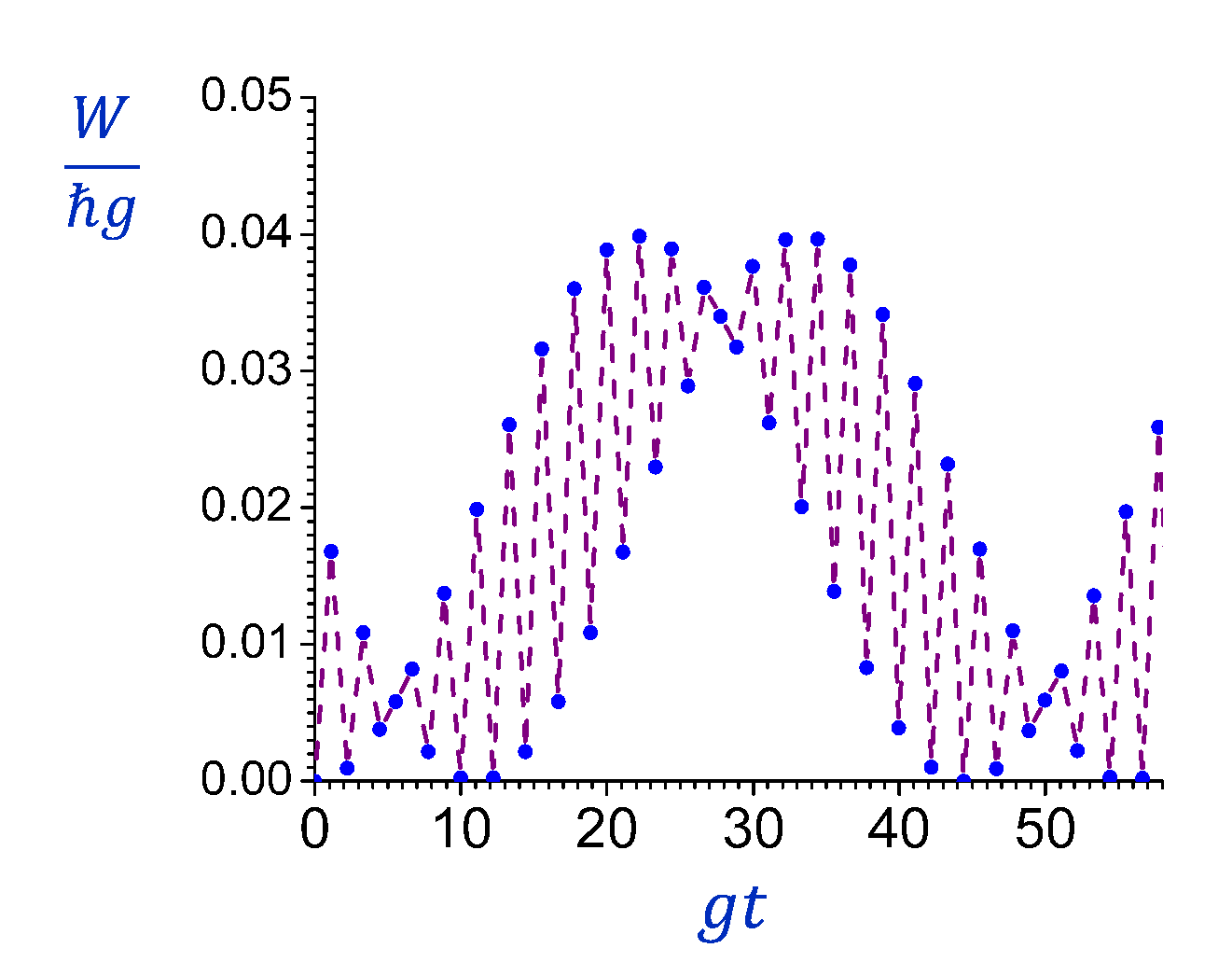}
\caption{Work done by photonic QHE as a function of the engine cycle
duration obtained for $\protect\omega_a-\protect\omega_b=2g$, $\protect%
\delta =0.2g$, and $\protect\nu =\protect\sqrt{4g^{2}+(\protect\omega _{a}-%
\protect\omega _{b})^{2}}=2\protect\sqrt{2}g$. The plot consists of discrete
points because in order to complete the cycle we must stop the system
evolution when oscillator frequencies return to their initial values. For
visualization we connected dots by a dash line. The maximum work is achieved
when the oscillator excitation numbers are exchanged.}
\label{WD}
\end{figure}

To close the cycle the system's evolution should be terminated when the
cavity frequencies return to their initial values. Then, we obtain a
discrete plot of $W$ as a function of time moments for which condition 
\begin{equation}
\omega _{a}(t)=\omega _{a}(0),\quad \omega _{b}(t)=\omega _{b}(0)  \label{K1}
\end{equation}
is satisfied (see Fig. \ref{WD}).

\begin{figure}[htbp]
\centering
\includegraphics[width=1\columnwidth]{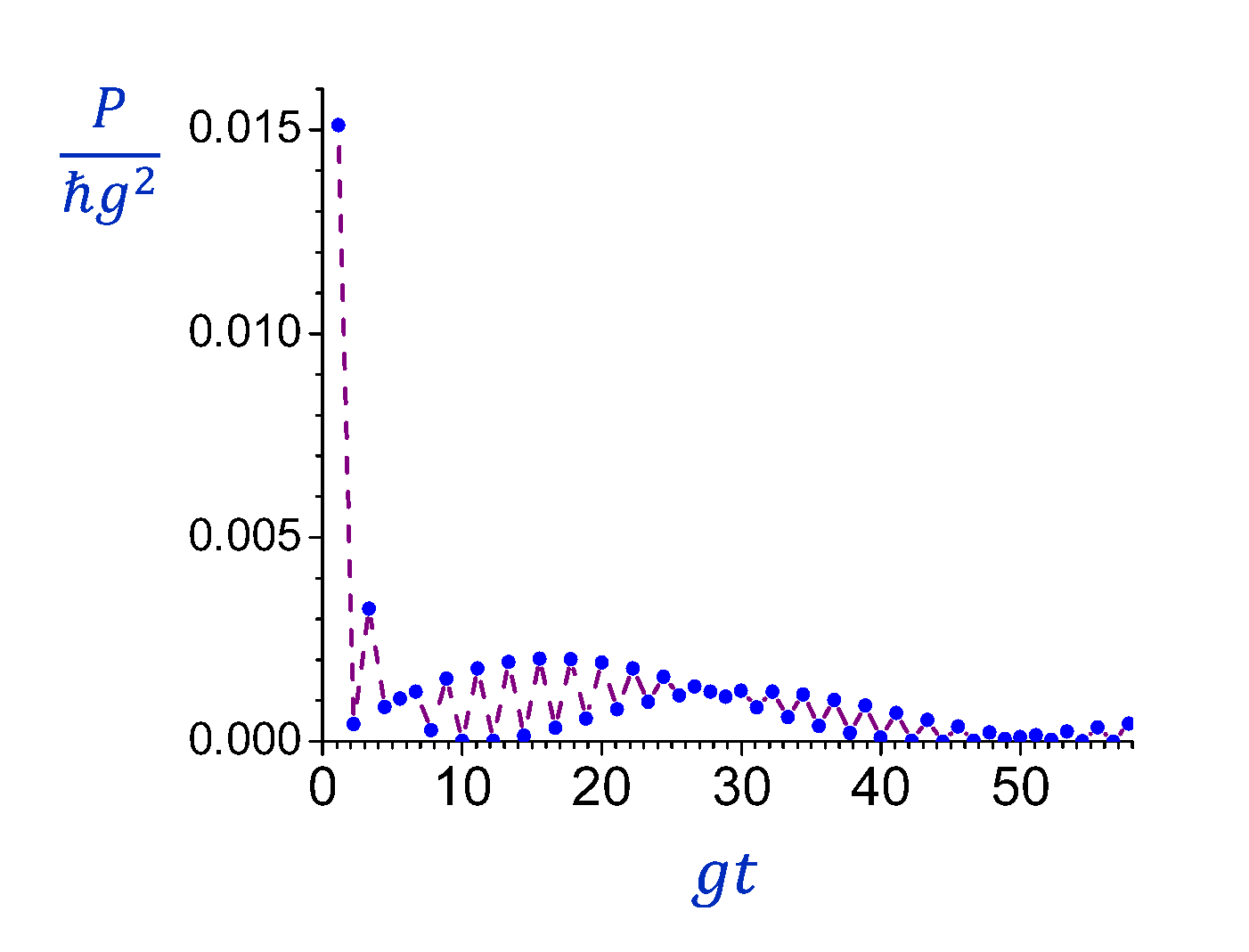}
\caption{Photonic QHE average power $P(t)=W(t)/t$ as a function of the
engine cycle duration obtained for parameters of Fig. \protect\ref{WD}. The
plot is a set of discrete points which we connected with dash line for
visualization. The engine operates at maximum power if the cycle is
terminated at the first instance the frequencies return to their initial
values. }
\label{AvgPD}
\end{figure}

Furthermore, we can calculate the average power per cycle $W(t)/t$ if we end
the cycle at different moments of time satisfying Eq. (\ref{K1}). It is
plotted in Fig. \ref{AvgPD}. Figure \ref{AvgPD} shows that engine power is
maximum if we stop the cycle at the first instance the frequencies are
restored to their initial values.

Finally, we note that Eq. (\ref{c2}) follows from the conservation of the
total number of excitations given by Eq. (\ref{c1}). If we choose
interaction Hamiltonian between oscillators as 
\begin{equation}
\hat{V}=\hbar g(a^{\dagger k}b^{m}+a^{k}b^{\dagger m}),
\end{equation}%
where $k$ and $m$ are positive integers, then Eqs. (\ref{c1}) and (\ref{c2})
are no longer valid. Instead, the combination%
\begin{equation}
kn_{a}(t)+mn_{b}(t)=kn_{a}(0)+mn_{b}(0)  \label{C4}
\end{equation}%
is now conserved. Using Eq. (\ref{C4}) we find that work per cycle is given
by 
\begin{equation*}
W=\hbar \left( \omega _{a}-\frac{k}{m}\omega _{b}\right)
[n_{a}(0)-n_{a}(t_{c})]
\end{equation*}%
and for the engine efficiency we obtain%
\begin{equation*}
\eta =\frac{W}{Q}=1-\frac{k\omega _{b}}{m\omega _{a}}.
\end{equation*}

\section{Atomic quantum heat engine}

\label{Sec5}

In this section, we consider a QHE whose working substance is made of
coupled two-level atoms with time-dependent transition frequencies $\omega
_{a}$ and $\omega _{b}$. The system's Hamiltonian reads%
\begin{equation}
\hat{H}=\frac{1}{2}\hslash \omega _{a}(t)\hat{\sigma}_{z}+\frac{1}{2}\hslash
\omega _{b}(t)\hat{\pi}_{z}+\hslash g\left( \hat{\sigma}^{\dag }\hat{\pi}+%
\hat{\sigma}\hat{\pi}^{\dag }\right) ,  \label{q1b}
\end{equation}%
where atomic operators obey commutation relations (\ref{CR1}).

We assume that the atomic QHE operates following the same cycle as the
photonic QHE of the previous section. Namely, at the beginning of the cycle
the atoms are not coupled with each other and are in contact with hot and
cold thermal reservoirs at temperatures $T_{h}$ and $T_{c}$. Then, atoms are
disconnected from the reservoirs and become coupled with each other. During
the engine operation, atomic transition frequencies change, e.g., by means
of the Stark or Zeeman effect and the system does the mechanical work on the
surroundings.

Since the operator $\hat{\sigma}_{z}+\hat{\pi}_{z}$ commutes with the
Hamiltonian, the total average number of excitations is conserved during the
system's evolution%
\begin{equation}
n_{a}(t)+n_{b}(t)=n_{a}(0)+n_{b}(0),  \label{q2a}
\end{equation}%
where%
\begin{equation}
n_{a}(t)=\frac{1}{2}+\frac{1}{2}\text{Tr}\left( \hat{\sigma}_{z}\hat{\rho}%
(t)\right) ,  \label{s1}
\end{equation}%
\begin{equation}
n_{b}(t)=\frac{1}{2}+\frac{1}{2}\text{Tr}\left( \hat{\pi}_{z}\hat{\rho}%
(t)\right)  \label{s2}
\end{equation}%
are the average numbers of excitations of the atom $a$ and $b$ respectively.
Hence, the efficiency of the engine is given by the same Eq. (\ref{c2}) as
in the previous section

\begin{equation}
\eta =1-\frac{\omega _{b}}{\omega _{a}}.  \label{p21}
\end{equation}

However, the work done by the engine is positive not for all values of $%
\omega _{a}$ and $\omega _{b}$. To find the sign of $W,$ \ we will use the
interaction picture in which the interaction Hamiltonian reads%
\begin{equation}
\hat{V}=\hslash g\left( e^{i\varphi (t)}\hat{\sigma}^{\dag }\hat{\pi}%
+e^{-i\varphi (t)}\hat{\sigma}\hat{\pi}^{\dag }\right) ,
\end{equation}%
where 
\begin{equation*}
\varphi (t)=\int_{0}^{t}dt^{\prime }(\omega _{b}(t^{\prime })-\omega
_{a}(t)).
\end{equation*}

Straightforward calculations yield the following equation for $n_{a}(t)$
(see Appendix B)%
\begin{equation*}
\left( \frac{\partial n_{a}}{\partial t}\right) ^{2}+\left( \int_{0}^{t}dt%
\frac{\partial \varphi }{\partial t}\frac{\partial n_{a}}{\partial t}\right)
^{2}=
\end{equation*}%
\begin{equation}
4g^{2}\left( n_{a}(0)-n_{a}(t)\right) \left( n_{a}(t)-n_{b}(0)\right) .
\label{p20}
\end{equation}%
Since the left-hand-side of Eq. (\ref{p20}) is nonnegative, we obtain that
at the end of the cycle $(t=t_{c})$, $n_{a}$ satisfies inequalities%
\begin{equation*}
n_{b}(0)\leq n_{a}(t_{c})\leq n_{a}(0),
\end{equation*}%
or%
\begin{equation*}
n_{a}(0)\leq n_{a}(t_{c})\leq n_{b}(0).
\end{equation*}%
Thus, if 
\begin{equation}
n_{a}(0)>n_{b}(0)  \label{s4}
\end{equation}%
then $n_{a}(t_{c})\leq n_{a}(0)$ and, according to Eq. (\ref{c1a}), work
done by the system during one cycle is nonnegative.

For a two-level atom%
\begin{equation*}
n_{a}(0)=\frac{1}{e^{\frac{\hslash \omega _{a}}{k_{B}T_{h}}}+1},\quad
n_{b}(0)=\frac{1}{e^{\frac{\hslash \omega _{b}}{k_{B}T_{c}}}+1},
\end{equation*}%
and condition (\ref{s4}) yields $\omega _{b}/\omega _{a}>T_{c}/T_{h}$. Plug
this into Eq. (\ref{p21}), we obtain that maximum efficiency of the heat
engine is again given by the Carnot formula (\ref{s5}).

The atomic QHE can be constructed, e.g., using two atoms separated by a few
atomic size distance and interacting with each other via dipole-dipole
interaction. If population of the higher atomic excited states is negligible
the Hamiltonian of such system can be approximated by Eq. (\ref{q1b}), where 
$\hat{\sigma}$ and $\hat{\pi}$ are transition operators between the ground
and the first excited state of $a$ and $b$ atom respectively. The
dipole-dipole interaction allows energy transfer between the atoms via force
of Coulomb repulsion between electrons in different atoms. This process does
not involve photon exchange and is described by the last term in the
Hamiltonian (\ref{q1b}).

\begin{figure}[h]
\begin{center}
\epsfig{figure=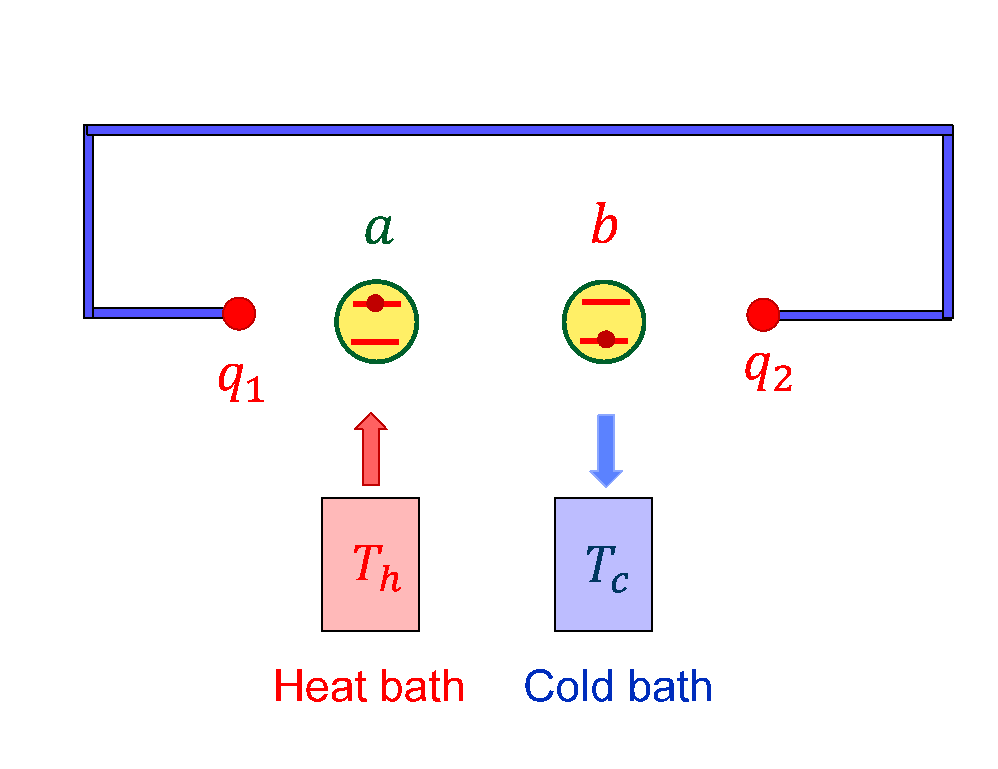, angle=0, width=8cm}
\end{center}
\caption{Atomic quantum heat engine composed of two atoms $a$ and $b$
coupled with each other via dipole-dipole interaction. Atomic transition
frequencies are varied by applying external AC electric field to the atoms
produced by moving electric charges $q_{1}$ and $q_{2}$. Atoms exert force
on the moving charges which do mechanical work on the surroundings.}
\label{AHE}
\end{figure}

One can make the atomic transition energy time dependent by applying
external AC electric or magnetic field at the location of the atoms. During
engine operation atoms do work on the field source. For example, AC external
electric field can be created by placing a moving electric charge near the
atom. Such charge plays a role of a piston. Moving the charge closer or
further from the atom produces work on the charge (see Fig. \ref{AHE}).

One should also add the working protocol. As long as we are able to maintain
the total excitation number conservation (that is keep photon emission
negligible), work is maximized if we pause the evolution when two atoms swap
their excitation numbers. After that we connect the atoms to hot and cold
baths respectively. Moreover, to close the cycle one needs to restore the
initial atomic frequencies at the end of the cycle.

\section{Quantum heat engine with counter-rotating interaction Hamiltonian}

\label{Sec6}

In this section, we discuss a QHE whose working substance is made of two
coupled harmonic oscillators of frequencies $\omega _{a}(t)$ and $\omega
_{b}(t)$ described by the Hamiltonian 
\begin{equation}
\hat{H}=\hslash \omega _{a}(t)\hat{a}^{\dag }\hat{a}+\hslash \omega _{b}(t)%
\hat{b}^{\dag }\hat{b}+\hslash g\left( \hat{a}^{\dag }\hat{b}^{\dag }+\hat{a}%
\hat{b}\right) .  \label{q1a}
\end{equation}%
The coupling between oscillators is governed by the counter-rotating
interaction term. The Hamiltonian with counter-rotating interaction appears,
e.g., in the theory of weakly interacting Bose-Einstein condensate in the
Bogoliubov approximation \cite{Pita03}.

For Hamiltonian (\ref{q1a}), the quantity $n_{a}(t)-n_{b}(t)$ is conserved
since operator $\hat{a}^{\dag }\hat{a}-\hat{b}^{\dag }\hat{b}$ commutes with 
$\hat{H}$. That is 
\begin{equation}
n_{a}(t)-n_{b}(t)=n_{a}(0)-n_{b}(0).  \label{s8}
\end{equation}%
In the interaction picture, the interaction Hamiltonian reads%
\begin{equation}
\hat{V}=\hslash g\left( e^{i\varphi (t)}\hat{a}^{\dag }\hat{b}^{\dag
}+e^{-i\varphi (t)}\hat{a}\hat{b}\right) ,
\end{equation}%
where 
\begin{equation*}
\varphi (t)=\int_{0}^{t}dt^{\prime }(\omega _{a}(t^{\prime })+\omega
_{b}(t^{\prime })).
\end{equation*}%
Straightforward derivation yields the following equation for $%
n_{a}(t)+n_{b}(t)$ (see Appendix C)

\begin{equation*}
\left[ \frac{\partial }{\partial t}\left( n_{a}+n_{b}\right) \right] ^{2}+%
\left[ \int_{0}^{t}dt\frac{\partial \varphi }{\partial t}\frac{\partial }{%
\partial t}\left( n_{a}+n_{b}\right) \right] ^{2}
\end{equation*}%
\begin{equation}
=4g^{2}\left( \left[ n_{a}+n_{b}+1\right] ^{2}-\left[ n_{a}(0)+n_{b}(0)+1%
\right] ^{2}\right) .  \label{s6}
\end{equation}%
Equation (\ref{s6}) shows that during engine operation 
\begin{equation}
n_{a}(t)+n_{b}(t)\geq n_{a}(0)+n_{b}(0).  \label{s7}
\end{equation}%
Combining this with Eq. (\ref{s8}) yields%
\begin{equation}
n_{a}(t)\geq n_{a}(0),\quad n_{b}(t)\geq n_{b}(0).  \label{s7a}
\end{equation}%
That is during system's evolution the numbers of oscillator excitations
increases. If the oscillators are not coupled at the beginning ($t=0$) and
the end ($t=t_{c}$) of the cycle, the work done by the system during the
cycle is%
\begin{equation}
W=\hslash \omega _{a}\left( n_{a}(0)-n_{a}(t_{c})\right) +\hslash \omega
_{b}\left( n_{b}(0)-n_{b}(t_{c})\right) .  \label{s7b}
\end{equation}%
Equations (\ref{s7a}) and (\ref{s7b}) yield that such QHE cannot produce
positive work for any closed cycle.

\section{Limiting efficiency of quantum heat engines}

\label{Sec7}

The second law of thermodynamics in the classical regime imposes a
fundamental limit on the maximum heat-to-work conversion efficiency in an
engine, given by Carnot efficiency. This efficiency is only achieved when
the engine operates in a cycle using reversible transformations, which
requires it to run infinitely slowly. As a consequence, the engine's power
(work extracted per unit time) becomes close to zero.

In this section, we consider a general case of a QHE whose working substance
is made of two systems with frequencies $\omega _{a}(t)$ and $\omega _{b}(t)$%
, and show that limiting efficiency of such engine is given by the Carnot
formula under the assumption that engine's power vanishes at the maximum
operating efficiency. We assume that at the beginning of the cycle the
working substance is in the thermal state (\ref{q2}) with $q=\hslash \omega
_{a}/k_{B}T_{h}$ and $p=\hslash \omega _{b}/k_{B}T_{c}$. We also assume that
during the engine operation the systems are isolated from the environment
but are coupled with each other. While at the beginning and the end of the
cycle the systems are connected with the hot and cold thermal reservoirs and
are not coupled with each other.

Since at the end of the cycle (at $t=t_{c}$) the frequencies return to their
initial values $\omega _{a}$ and $\omega _{b}$, the work done by the engine
during one cycle is given by Eq. (\ref{s7b}), while heat taken from the hot
reservoir per cycle is%
\begin{equation*}
Q=\hslash \omega _{a}\left( n_{a}(0)-n_{a}(t_{c})\right) .
\end{equation*}%
Hence, the engine efficiency is given by%
\begin{equation}
\eta =\frac{W}{Q}=1-\frac{\omega _{b}}{\omega _{a}}\frac{%
n_{b}(t_{c})-n_{b}(0)}{n_{a}(0)-n_{a}(t_{c})}.  \label{s10}
\end{equation}

If at the maximum efficiency operation the engine power vanishes the density
operator (\ref{q2}) does not evolve with time in this regime. The time
evolution of the density operator is given by%
\begin{equation*}
\hat{\rho}(t)=Ne^{-q\hat{n}_{a}(t)-p\hat{n}_{b}(t)},
\end{equation*}%
where $\hat{n}_{a}(t)=\hat{a}^{\dag }(t)\hat{a}(t)$ and $\hat{n}_{b}(t)=\hat{%
b}^{\dag }(t)\hat{b}(t)$. If $\hat{\rho}$ is independent of $t$ the quantity%
\begin{equation*}
qn_{a}(t)+pn_{b}(t)
\end{equation*}%
is conserved during the engine operation which yields%
\begin{equation*}
\frac{n_{b}(t_{c})-n_{b}(0)}{n_{a}(0)-n_{a}(t_{c})}=\frac{q}{p}.
\end{equation*}

Plug this in Eq. (\ref{s10}), and taking into account that $q/p=T_{c}\omega
_{a}/T_{h}\omega _{b}$, gives that the maximum engine efficiency is given by
the Carnot formula. The present analysis provides an alternative derivation
of the Carnot limit on QHE efficiency \cite{Pusz78,Alic79}.

\section{Quantum heat engines based on correlated Gaussian states}

\label{Gaussian}

Gaussian states are described by the density operator of the form

\begin{equation}
\hat{\rho}=Ne^{-H_{ij}\hat{a}_{i}^{\dag }\hat{a}_{j}+(K_{ij}\hat{a}%
_{i}^{\dag }\hat{a}_{j}^{\dag }+h.c.)},  \label{GG1}
\end{equation}%
where repeated indices are implicitly summed over; $i,j=1,\ldots ,n$.
Equation (\ref{GG1}) defines a general Gaussian state for $n$ modes, each of
which is described by the creation (annihilation) operator $\hat{a}%
_{i}^{\dag }$ ($\hat{a}_{i}$), which we assume obey bosonic commutation
relations. Without loss of generality, one can assume that $H_{ij}$ and $%
K_{ij}$ are real numbers. If not, the complex phases of $H_{ij}$ and $K_{ij}$
can be subsumed into operators $\hat{a}_{i}$ by replacing $\hat{a}%
_{i}\rightarrow \hat{a}_{i}e^{i\varphi _{i}}$.

Introducing a vector%
\begin{equation*}
\hat{a}=\left( 
\begin{array}{c}
\hat{a}_{1} \\ 
\vdots \\ 
\hat{a}_{n} \\ 
\hat{a}_{1}^{\dag } \\ 
\vdots \\ 
\hat{a}_{n}^{\dag }%
\end{array}%
\right) ,
\end{equation*}%
one can write $\hat{\rho}$ as 
\begin{equation}
\hat{\rho}=Ne^{-\hat{a}^{\dag }G\hat{a}},  \label{GG11}
\end{equation}%
where $G$ is $2n\times 2n$ real Hermitian matrix, and $\hat{a}^{\dag }$
denotes a vector%
\begin{equation*}
\hat{a}^{\dag }=\left( 
\begin{array}{cccccc}
\hat{a}_{1}^{\dag } & \ldots & \hat{a}_{n}^{\dag } & \hat{a}_{1} & \ldots & 
\hat{a}_{n}%
\end{array}%
\right) .
\end{equation*}

Next, we use the Williamson theorem \cite{Will36} which states that there
exists a $2n\times 2n$ real symplectic matrix $S$ such that

\begin{equation*}
G=S^{\dag }DS,
\end{equation*}%
where%
\begin{equation*}
D=\frac{1}{2}\text{diag}\left( \beta _{1},\ldots ,\beta _{n},\beta
_{1},\ldots ,\beta _{n}\right)
\end{equation*}%
is a real diagonal $2n\times 2n$ matrix. The symplectic matrix $S$ has the
block structure 
\begin{equation}
S=\left( 
\begin{array}{cc}
C & P \\ 
M & L%
\end{array}%
\right) ,
\end{equation}%
where $C,P,M,L$ are $n\times n$ real matrices satisfying the following
conditions%
\begin{equation}
CP^{T}=PC^{T},\qquad ML^{T}=LM^{T},  \label{GGQ1}
\end{equation}%
\begin{equation}
CL^{T}-PM^{T}=I.  \label{GGQ2}
\end{equation}%
Here $C^{T}$ stands for the transpose of the matrix $C$, and $I$ is a unit $%
n\times n$ matrix. Conditions (\ref{GGQ1}) and (\ref{GGQ2}) follow from the
definition of symplectic matrix. Namely, symplectic matrix is a $2n\times 2n$
real matrix $S$ satisfying the condition%
\begin{equation}
S^{T}\Omega S=\Omega ,  \label{GGQ3}
\end{equation}%
where 
\begin{equation}
\Omega =\left( 
\begin{array}{cc}
0 & I \\ 
-I & 0%
\end{array}%
\right) .  \label{GGQ4}
\end{equation}

Next, we introduce operators $\hat{A}_{i}$ ($i=1,\ldots ,n$) according to
the equation (symplectic transformation)%
\begin{equation*}
\hat{A}=S\hat{a},\qquad \hat{A}^{\dag }=\hat{a}^{\dag }S^{\dag },
\end{equation*}%
where%
\begin{equation*}
\hat{A}=\left( 
\begin{array}{c}
\hat{A}_{1} \\ 
\vdots \\ 
\hat{A}_{n} \\ 
\hat{A}_{1}^{\dag } \\ 
\vdots \\ 
\hat{A}_{n}^{\dag }%
\end{array}%
\right) .
\end{equation*}

Operators $\hat{A}_{k}$ are linear combinations of $\hat{a}_{i}$ and $\hat{a}%
_{i}^{\dag }$, which describe collective excitations of the oscillators $%
\hat{a}_{i}$. They obey the same bosonic commutation relations as the
operators $\hat{a}_{i}$. Indeed,%
\begin{equation*}
\lbrack \hat{A}_{i},\hat{A}_{j}^{\dag }]=[C_{ik}\hat{a}_{k}+P_{ik}\hat{a}%
_{k}^{\dag },M_{jm}\hat{a}_{m}+L_{jm}\hat{a}_{m}^{\dag }]=
\end{equation*}%
\begin{equation*}
=C_{ik}L_{jk}-P_{ik}M_{jk}=\left( CL^{T}-PM^{T}\right) _{ij}=\delta _{ij},
\end{equation*}%
where we used the property (\ref{GGQ2}) of the symplectic matrix $S$.
Similarly,%
\begin{equation*}
\lbrack \hat{A}_{i},\hat{A}_{j}]=\left( CP^{T}-PC^{T}\right) _{ij}=0,
\end{equation*}%
\begin{equation*}
\lbrack \hat{A}_{i}^{\dag },\hat{A}_{j}^{\dag }]=\left( ML^{T}-LM^{T}\right)
_{ij}=0
\end{equation*}%
according to Eqs. (\ref{GGQ1}).

One can choose collective modes $\hat{A}_{i}$ as a new basis set to describe
the system. In terms of $\hat{A}_{i},$ the density operator of the Gaussian
state (\ref{GG11}) is a product of thermal density operators for each
collective mode $\hat{A}_{i}$ 
\begin{equation}
\hat{\rho}=Ne^{-\hat{a}^{\dag }G\hat{a}}=\tilde{N}e^{-\beta _{1}\hat{A}%
_{1}^{\dag }\hat{A}_{1}-\beta _{2}\hat{A}_{2}^{\dag }\hat{A}_{2}-\ldots
-\beta _{n}\hat{A}_{n}^{\dag }\hat{A}_{n}},  \label{GG10}
\end{equation}%
where the normalization factor is%
\begin{equation*}
\tilde{N}=\prod\limits_{i=1}^{n}\left( 1-e^{-\beta _{i}}\right) .
\end{equation*}

One can also factorize Gaussian density operator for fermions. In this case,
instead of the symplectic matrix, one should use matrix $S$ obeying Eq. (\ref%
{GGQ3}), but with 
\begin{equation}
\Omega =\left( 
\begin{array}{cc}
0 & I \\ 
I & 0%
\end{array}%
\right) ,
\end{equation}%
which guarantees that collective excitation modes obey the same
anti-commutation relations for the fermionic operators. This yields that for
fermions the matrix $S$ is unitary ($S^{\dag }S=SS^{\dag }=I_{2n}$) with the
block structure%
\begin{equation*}
S=\left( 
\begin{array}{cc}
C & P \\ 
P^{\ast } & C^{\ast }%
\end{array}%
\right) ,
\end{equation*}%
and 
\begin{equation*}
G=S^{\dag }DS,
\end{equation*}%
where%
\begin{equation*}
D=\frac{1}{2}\text{diag}\left( \beta _{1},\ldots ,\beta _{n},-\beta
_{1},\ldots ,-\beta _{n}\right) .
\end{equation*}

Obviously, the same procedure can be used to diagonalize arbitrary quadratic
boson and fermion\textit{\ }Hamiltonians in terms of quasiparticles (for a
simple proof see Refs. \cite{Hemm80,Nica21}), which generalizes Bogoliubov
transformations. The latter are widely used in description of
superconductivity, superfluidity, spin waves in ferromagnets, etc.

We obtain that, physically, the state of the system (\ref{GG1}) is
equivalent to that composed of $n$ independent thermal modes at different
temperatures. These collective thermal modes can be used as a heat and cold
baths in a QHE. That is, operation of the QHE based on the Gaussian states
is equivalent to that based on uncorrelated thermal reservoirs. In
particular, for a two-mode system, the maximum engine efficiency will be
given by the Carnot formula in which $T_{h}$ and $T_{c}$ are the
temperatures of the corresponding collective thermal modes in Eq. (\ref{GG10}%
). In terms of the original modes $\hat{a}_{j},$ the engine operation might
look like work production from a single correlated reservoir.

As an example, we consider a two-mode system described by the bosonic
operators $\hat{a}$ and $\hat{b},$ prepared in a correlated Gaussian state%
\begin{equation}
\hat{\rho}=Ne^{-q(\hat{a}^{\dag }\hat{a}+\hat{b}^{\dag }\hat{b})+i\gamma (%
\hat{a}^{\dag }\hat{b}-\hat{a}\hat{b}^{\dag })},  \label{GG2}
\end{equation}%
where $\gamma >0$ and $N$ is the normalization factor%
\begin{equation*}
N=\left( 1-e^{-q-\gamma }\right) \left( 1-e^{-q+\gamma }\right) .
\end{equation*}%
In terms of the canonical collective modes%
\begin{equation*}
\hat{A}=\frac{\hat{a}+i\hat{b}}{\sqrt{2}},\qquad \hat{B}=\frac{\hat{b}+i\hat{%
a}}{\sqrt{2}},
\end{equation*}%
the state (\ref{GG2}) is a product of two thermal states with different
temperatures%
\begin{equation}
\hat{\rho}=Ne^{-(q-\gamma )\hat{A}^{\dag }\hat{A}-(q+\gamma )\hat{B}^{\dag }%
\hat{B}}.  \label{GG3}
\end{equation}

Thus, the maximum efficiency of a QHE operating based on the state (\ref{GG2}%
) is given by 
\begin{equation}
\eta _{\max }=1-\frac{q-\gamma }{q+\gamma }=\frac{2\gamma }{q+\gamma }.
\label{s5GG}
\end{equation}%
That is, $\eta _{\max }$ is determined by the correlation strength $\gamma $.

If the system is prepared in a two-mode squeezed thermal state%
\begin{equation}
\hat{\rho}=Ne^{-q(\hat{a}^{\dag }\hat{a}+\hat{b}^{\dag }\hat{b})+\gamma (%
\hat{a}^{\dag }\hat{b}^{\dag }+\hat{a}\hat{b})},  \label{GG12}
\end{equation}%
then in terms of the collective (Bogoliubov) modes%
\begin{equation*}
\hat{A}=\frac{\hat{a}-\mu \hat{b}^{\dag }}{\sqrt{1-\mu ^{2}}},\qquad \hat{B}=%
\frac{\hat{b}-\mu \hat{a}^{\dag }}{\sqrt{1-\mu ^{2}}},
\end{equation*}%
where%
\begin{equation}
\mu =\frac{1}{\gamma }\left( q-\sqrt{q^{2}-\gamma ^{2}}\right) ,  \label{mu1}
\end{equation}%
the state (\ref{GG12}) is a product of two thermal states with equal
temperatures%
\begin{equation}
\hat{\rho}=\left( 1-e^{-\sqrt{q^{2}-\gamma ^{2}}}\right) ^{2}e^{-\sqrt{%
q^{2}-\gamma ^{2}}\left( \hat{A}^{\dag }\hat{A}+\hat{B}^{\dag }\hat{B}%
\right) }.
\end{equation}%
That is, squeezing effectively increases temperature of each oscillator by
an equal amount. However, since temperatures remain equal, no work can be
produced by the single squeezed reservoir in a closed cycle.

Similar increase of the effective temperature takes place for a single-mode
squeezed thermal state%
\begin{equation}
\hat{\rho}=Ne^{-q\hat{a}^{\dag }\hat{a}+\frac{\gamma }{2}(\hat{a}^{\dag }%
\hat{a}^{\dag }+\hat{a}\hat{a})},
\end{equation}%
which, in terms of the canonical Bogoliubov mode 
\begin{equation*}
\hat{A}=\frac{\hat{a}-\mu \hat{a}^{\dag }}{\sqrt{1-\mu ^{2}}},\qquad \hat{A}%
^{\dag }=\frac{\hat{a}^{\dag }-\mu \hat{a}}{\sqrt{1-\mu ^{2}}},
\end{equation*}%
where $\mu $ is given by Eq. (\ref{mu1}), reduces to a thermal state with
higher temperature%
\begin{equation}
\hat{\rho}=\left( 1-e^{-\sqrt{q^{2}-\gamma ^{2}}}\right) e^{-\sqrt{%
q^{2}-\gamma ^{2}}\hat{A}^{\dag }\hat{A}}.
\end{equation}

A more general two-mode Gaussian state 
\begin{equation}
\hat{\rho}=Ne^{-q(\hat{a}^{\dag }\hat{a}+\hat{b}^{\dag }\hat{b})+i\gamma
_{1}(\hat{a}^{\dag }\hat{b}-\hat{a}\hat{b}^{\dag })-i\gamma _{2}(\hat{a}%
^{\dag }\hat{b}^{\dag }-\hat{a}\hat{b})},
\end{equation}%
in terms of the collective operators 
\begin{equation*}
\hat{A}=\frac{\hat{a}+i\hat{b}-\mu _{-}\hat{a}^{\dag }+i\mu _{-}\hat{b}%
^{\dag }}{\sqrt{2}\sqrt{1-\mu _{-}^{2}}},
\end{equation*}%
\begin{equation*}
\hat{B}=\frac{\hat{b}+i\hat{a}-\mu _{+}\hat{b}^{\dag }+i\mu _{+}\hat{a}%
^{\dag }}{\sqrt{2}\sqrt{1-\mu _{+}^{2}}},
\end{equation*}%
obeying bosonic commutation relations, where%
\begin{equation*}
\mu _{\pm }=\frac{1}{\gamma _{2}}\left( q\pm \gamma _{1}-\sqrt{\left( q\pm
\gamma _{1}\right) ^{2}-\gamma _{2}^{2}}\right) ,
\end{equation*}%
reduces to a product of two thermal states with different temperatures%
\begin{equation}
\hat{\rho}=\left( 1-e^{-\beta_{-}}\right) \left( 1-e^{-\beta_{+}}\right)
e^{-\beta_{-}\hat{A}^{\dag }\hat{A}-\beta_{+}\hat{B}^{\dag }\hat{B}},
\end{equation}%
with%
\begin{equation*}
\beta_{\pm }=\sqrt{\left( q\pm \gamma _{1}\right) ^{2}-\gamma _{2}^{2}}.
\end{equation*}

According to the procedure developed in the present paper, evolution of the
density operator (\ref{GG1}) under the Hamiltonian $\hat{H}$ is obtained by
replacing operators $\hat{a}_{j}$ in Eq. (\ref{GG1}) with operators $\hat{a}%
_{j}(t)$. The latter obey differential equations%
\begin{equation*}
i\hslash \frac{\partial \hat{a}_{j}(t)}{\partial t}=\left[ \hat{H},\hat{a}%
_{j}(t)\right] ,\qquad j=1,\ldots ,n,
\end{equation*}%
subject to the initial conditions $\hat{a}_{j}(0)=\hat{a}_{j}.$ In
particular, the two-mode state (\ref{GG2}), under the Hamiltonian%
\begin{equation*}
\hat{H}=\hslash \omega \left( \hat{a}^{\dag }\hat{a}+\hat{b}^{\dag }\hat{b}%
\right) +\hslash g\left( \hat{a}^{\dag }\hat{b}+\hat{a}\hat{b}^{\dag
}\right) ,
\end{equation*}%
evolves as%
\begin{equation*}
\hat{\rho}(t)=Ne^{-(q+\gamma \sin (2gt))\hat{a}^{\dag }\hat{a}-(q-\gamma
\sin (2gt))\hat{b}^{\dag }\hat{b}+i\gamma \cos (2gt)(\hat{a}^{\dag }\hat{b}-%
\hat{a}\hat{b}^{\dag })}.
\end{equation*}%
For the time instances for which $\sin (2gt)=\pm 1,$ the state becomes a
product of thermal states with different temperatures%
\begin{equation*}
\hat{\rho}=Ne^{-(q\pm \gamma )\hat{a}^{\dag }\hat{a}-(q\mp \gamma )\hat{b}%
^{\dag }\hat{b}}.
\end{equation*}

\section{Summary}

In this paper, we introduce a technique for calculation the density operator
time evolution along the lines of Heisenberg representation of quantum
mechanics. The technique is applicable when initially the system is prepared
in a mixed state but subsequent evolution is unitary, that is during the
evolution the system is isolated from the environment. Under these
conditions, the initial density operator 
\begin{equation}
\hat{\rho}_{0}=\hat{\rho}_{0}(\hat{a},\hat{b},\ldots ),
\end{equation}%
where $\hat{a},$ $\hat{b},$ $\ldots $ are operators describing the system,
evolves as 
\begin{equation}
\hat{\rho}(t)=\hat{\rho}_{0}(\hat{A}(t),\hat{B}(t),\ldots ),
\end{equation}%
where operators $\hat{A}(t)$, $\hat{B}(t)$, \ldots\ obey the same equation
of motion as the density operator, namely%
\begin{equation}
i\hslash \frac{\partial \hat{A}(t)}{\partial t}=\left[ \hat{H},\hat{A}(t)%
\right] ,\quad \ldots
\end{equation}%
and subject to the initial conditions $\hat{A}(0)=\hat{a},$ $\hat{B}(0)=\hat{%
b}$, etc.

Using this formalism, we find an exact solution for the evolution of two and
three coupled harmonic oscillators as well as two-level atoms initially
prepared in thermal states at different temperatures. We show that such
systems exhibit interesting quantum dynamics in which oscillators swap their
thermal states (temperatures) due to correlations induced in the process of
energy exchange and yield noise induced coherence. We also show that if
initially oscillators are in a coherent state then during evolution all
population of one oscillator can be transferred to another oscillator.

In addition, we investigate QHEs that use hot and cold reservoirs in thermal
equilibrium as the energy source and the entropy sink respectively. We model
a working substance as coupled harmonic oscillators (photonic QHE) or
coupled two-level systems (atomic QHE) whose transition frequencies depend
on time. We show that the efficiency of such engines depends only on the
initial frequencies and is independent of how oscillator frequencies vary
during the engine operation. Work done by the engine per cycle is maximum if
at the end of the cycle the oscillators swap numbers of excitations which
can be achieved when the engine operates under the condition of parametric
resonance. In contrast, we show that the engine operates at maximum power if
the cycle is terminated at the first instance the frequencies return to
their initial values.

We also find that working substance composed of oscillators coupled only by
counter-rotating terms (BEC-like) is unable to produce positive work in any
closed cycle.

Finally, we show that Carnot formula yields the limiting efficiency for QHEs
under general assumptions, which provides an alternative way to derive the
Carnot limit on QHE efficiency \cite{Pusz78,Alic79}. Moreover, Carnot
formula with effective temperatures can be used to find limiting efficiency
for QHEs that use correlated energy reservoirs, as we discuss in Sect. \ref%
{Gaussian}.

The present technique can be applied for study the other problems involving
quantum evolution of mixed states, e.g., an atom uniformly accelerated
through Minkowski vacuum \cite{Svid25}. In terms of Rindler photons,
Minkowski vacuum is in a thermal state \cite{Unru84,Svid21}. That is to say
that we have an isolated system which is initially in a mixed state.

Our findings deepen understanding of quantum evolution of mixed states which
could be useful for design of quantum machines with better performance.

\begin{acknowledgements}
This work was supported by U.S. Department of Energy (DE-SC-0023103; DE-AC36-08GO28308, SUB-2023-10388; DE-SC0024882); Welch Foundation (A-1261); Air Force Office of Scientific Research (FA9550-20-1-0366). 
W.Z. is supported by the Herman F. Heep and Minnie Belle Heep Texas A\&M University Endowed 
Fund held and administered by the Texas A\&M Foundation.
\end{acknowledgements}

\appendix

\section{Derivation of Eq. (\protect\ref{q3b})}

Here we show that equation 
\begin{equation}
\hat{\rho}(t)=\hat{U}^{\dagger }(t)\hat{\rho}_{0}\hat{U}(t)=Ne^{-q\hat{A}%
^{\dagger }(t)\hat{A}(t)}e^{-p\hat{B}^{\dagger }(t)\hat{B}(t)},  \label{W4}
\end{equation}%
where%
\begin{equation*}
\hat{\rho}_{0}\,(\hat{a},\hat{b})=Ne^{-q\hat{a}^{\dagger }\hat{a}}e^{-p\hat{b%
}^{\dagger }\hat{b}},
\end{equation*}%
\begin{equation*}
\hat{A}(t)=\hat{U}^{\dagger }(t)\,\hat{a}\,\hat{U}(t),\quad \hat{B}(t)=\hat{U%
}^{\dagger }(t)\,\hat{b}\,\hat{U}(t)
\end{equation*}
is exact as long as the evolution is unitary. To prove that we start from a
simple example, namely we calculate%
\begin{equation*}
\hat{U}^{\dagger }\,e^{-q\hat{a}^{\dagger }\hat{a}}\hat{U},
\end{equation*}
where $\hat{U}$ is a unitary operator 
\begin{equation}
\hat{U}\hat{U}^{\dagger }=\hat{U}^{\dagger }\hat{U}=1.  \label{W2}
\end{equation}

Making the Taylor expansion of the exponential%
\begin{equation*}
e^{-q\hat{a}^{\dagger }\hat{a}}=1-q\hat{a}^{\dagger }\hat{a}+\frac{q^{2}}{2}%
\hat{a}^{\dagger }\hat{a}\hat{a}^{\dagger }\hat{a}-\ldots
\end{equation*}
and using Eq. (\ref{W2}), we have%
\begin{equation*}
\hat{U}^{\dagger }\,e^{-q\hat{a}^{\dagger }\hat{a}}\hat{U}=\hat{U}^{\dagger }%
\hat{U}-q\hat{U}^{\dagger }\hat{a}^{\dagger }\hat{a}\hat{U}+\frac{q^{2}}{2}%
\hat{U}^{\dagger }\hat{a}^{\dagger }\hat{a}\hat{a}^{\dagger }\hat{a}\hat{U}%
-\ldots
\end{equation*}%
\begin{equation*}
=1-q\hat{U}^{\dagger }\hat{a}^{\dagger }\hat{U}\hat{U}^{\dagger }\hat{a}\hat{%
U}+\frac{q^{2}}{2}\hat{U}^{\dagger }\hat{a}^{\dagger }\hat{U}\hat{U}%
^{\dagger }\hat{a}\hat{U}\hat{U}^{\dagger }\hat{a}^{\dagger }\hat{U}\hat{U}%
^{\dagger }\hat{a}\hat{U}-\ldots
\end{equation*}%
\begin{equation*}
=1-q\hat{A}^{\dagger }\hat{A}+\frac{q^{2}}{2}\hat{A}^{\dagger }\hat{A}\hat{A}%
^{\dagger }\hat{A}-\ldots =e^{-q\hat{A}^{\dagger }\hat{A}},
\end{equation*}%
where we introduced operators%
\begin{equation*}
\hat{A}=\hat{U}^{\dagger }\,\hat{a}\,\hat{U},\quad \hat{A}^{\dagger }=\hat{U}%
^{\dagger }\hat{a}^{\dagger }\,\hat{U}.
\end{equation*}

Therefore%
\begin{equation*}
\hat{U}^{\dagger }\,e^{-q\hat{a}^{\dagger }\hat{a}}\hat{U}=e^{-q\hat{A}%
^{\dagger }\hat{A}}.
\end{equation*}

Next we calculate%
\begin{equation*}
\hat{U}^{\dagger }\,e^{-q\hat{a}^{\dagger }\hat{a}-p\hat{b}^{\dagger }\hat{b}%
}\hat{U}=\hat{U}^{\dagger }\,e^{-q\hat{a}^{\dagger }\hat{a}}\hat{U}\hat{U}%
^{\dagger }e^{-p\hat{b}^{\dagger }\hat{b}}\hat{U}
\end{equation*}%
\begin{equation*}
=e^{-q\hat{A}^{\dagger }\hat{A}}e^{-q\hat{B}^{\dagger }\hat{B}}=e^{-q\hat{A}%
^{\dagger }\hat{A}-q\hat{B}^{\dagger }\hat{B}},
\end{equation*}
where%
\begin{equation*}
\hat{B}=\hat{U}^{\dagger }\,\hat{b}\,\hat{U},\quad \hat{B}^{\dagger }=\hat{U}%
^{\dagger }\hat{b}^{\dagger }\,\hat{U}.
\end{equation*}%
This proves Eq. (\ref{W4}).

\section{Derivation of Eq. (\protect\ref{p20})}

Taking the time derivative of Eqs. (\ref{s1}) and (\ref{s2}), using the
evolution equation for the density operator 
\begin{equation}
i\hslash \frac{\partial \hat{\rho}}{\partial t}=\left[ \hat{V},\hat{\rho}%
\right]  \label{q2b}
\end{equation}%
and the invariance of trace under cyclic permutations, we have 
\begin{equation*}
i\hslash \frac{\partial n_{a}}{\partial t}=\frac{1}{2}\text{Tr}\left( [\hat{%
\sigma}_{z},\hat{V}]\hat{\rho}(t)\right)
\end{equation*}%
\begin{equation*}
=\hslash g\text{Tr}\left( \left( e^{i\varphi (t)}\hat{\sigma}^{\dag }\hat{\pi%
}-e^{-i\varphi (t)}\hat{\sigma}\hat{\pi}^{\dag }\right) \hat{\rho}(t)\right)
,
\end{equation*}%
\begin{equation*}
i\hslash \frac{\partial n_{b}}{\partial t}=\frac{1}{2}\text{Tr}\left( [\hat{%
\pi}_{z},\hat{V}]\hat{\rho}(t)\right)
\end{equation*}%
\begin{equation*}
=-\hslash g\text{Tr}\left( \left( e^{i\varphi (t)}\hat{\sigma}^{\dag }\hat{%
\pi}-e^{-i\varphi (t)}\hat{\sigma}\hat{\pi}^{\dag }\right) \hat{\rho}%
(t)\right) ,
\end{equation*}%
or%
\begin{equation*}
-\hslash \frac{\partial }{\partial t}\left( n_{a}-n_{b}\right) =2i\hslash g%
\text{Tr}\left( \left( e^{i\varphi (t)}\hat{\sigma}^{\dag }\hat{\pi}%
-e^{-i\varphi (t)}\hat{\sigma}\hat{\pi}^{\dag }\right) \hat{\rho}(t)\right) .
\end{equation*}%
Taking the time derivative again, we find%
\begin{equation*}
-\hslash \frac{\partial ^{2}}{\partial t^{2}}\left( n_{a}-n_{b}\right) =
\end{equation*}%
\begin{equation*}
-2\hslash g\dot{\varphi}\text{Tr}\left( \left( e^{i\varphi (t)}\hat{\sigma}%
^{\dag }\hat{\pi}+e^{-i\varphi (t)}\hat{\sigma}\hat{\pi}^{\dag }\right) \hat{%
\rho}(t)\right)
\end{equation*}%
\begin{equation*}
+2g\text{Tr}\left( \left[ e^{i\varphi (t)}\hat{\sigma}^{\dag }\hat{\pi}%
-e^{-i\varphi (t)}\hat{\sigma}\hat{\pi}^{\dag },\hat{V}\right] \hat{\rho}%
\right) =
\end{equation*}%
\begin{equation*}
-2\hslash g\dot{\varphi}\text{Tr}\left( \left( e^{i\varphi (t)}\hat{\sigma}%
^{\dag }\hat{\pi}+e^{-i\varphi (t)}\hat{\sigma}\hat{\pi}^{\dag }\right) \hat{%
\rho}(t)\right)
\end{equation*}%
\begin{equation*}
+2\hslash g^{2}\text{Tr}\left( \left( \left[ \hat{\sigma}^{\dag }\hat{\pi},%
\hat{\sigma}\hat{\pi}^{\dag }\right] -\left[ \hat{\sigma}\hat{\pi}^{\dag },%
\hat{\sigma}^{\dag }\hat{\pi}\right] \right) \hat{\rho}\right) =
\end{equation*}%
\begin{equation*}
-2\hslash g\dot{\varphi}\text{Tr}\left( \left( e^{i\varphi (t)}\hat{\sigma}%
^{\dag }\hat{\pi}+e^{-i\varphi (t)}\hat{\sigma}\hat{\pi}^{\dag }\right) \hat{%
\rho}(t)\right)
\end{equation*}%
\begin{equation*}
+2\hslash g^{2}\text{Tr}\left( \left( \hat{\sigma}_{z}-\hat{\pi}_{z}\right) 
\hat{\rho}\right) =-2\dot{\varphi}V(t)+4\hslash g^{2}\left(
n_{a}-n_{b}\right) ,
\end{equation*}%
where%
\begin{equation*}
V(t)=\hslash g\text{Tr}\left( \left( e^{i\varphi (t)}\hat{\sigma}^{\dag }%
\hat{\pi}+e^{-i\varphi (t)}\hat{\sigma}\hat{\pi}^{\dag }\right) \hat{\rho}%
(t)\right) .
\end{equation*}

Thus, we obtain a differential equation%
\begin{equation}
\frac{\partial ^{2}}{\partial t^{2}}\left( n_{a}-n_{b}\right) -\frac{2\dot{%
\varphi}}{\hslash }V(t)+4g^{2}\left( n_{a}-n_{b}\right) =0.  \label{s3}
\end{equation}%
Taking the time derivative of $V(t)$, we have%
\begin{equation*}
-\frac{2}{\hslash }\frac{\partial V(t)}{\partial t}=\dot{\varphi}\frac{%
\partial }{\partial t}\left( n_{a}-n_{b}\right) ,
\end{equation*}%
or%
\begin{equation*}
-\frac{2}{\hslash }V(t)=\int_{0}^{t}dt\dot{\varphi}\frac{\partial }{\partial
t}\left( n_{a}-n_{b}\right) .
\end{equation*}%
Therefore, Eq. (\ref{s3}) reduces to 
\begin{equation*}
\frac{\partial ^{2}}{\partial t^{2}}\left( n_{a}-n_{b}\right) +\dot{\varphi}%
\int_{0}^{t}dt\dot{\varphi}\frac{\partial }{\partial t}\left(
n_{a}-n_{b}\right) +4g^{2}\left( n_{a}-n_{b}\right) =0.
\end{equation*}%
Multiplying this equation by $\frac{\partial }{\partial t}\left(
n_{a}-n_{b}\right) $ and integrating over time, we find%
\begin{equation*}
\left[ \frac{\partial }{\partial t}\left( n_{a}-n_{b}\right) \right] ^{2}+%
\left[ \int_{0}^{t}dt\dot{\varphi}\frac{\partial }{\partial t}\left(
n_{a}-n_{b}\right) \right] ^{2}
\end{equation*}%
\begin{equation}
+4g^{2}\left( n_{a}-n_{b}\right) ^{2}=const.  \label{k1}
\end{equation}%
Taking into account Eq. (\ref{q2a}), one can write Eq. (\ref{k1}) as%
\begin{equation*}
\left( \frac{\partial n_{a}}{\partial t}\right) ^{2}+\left( \int_{0}^{t}dt%
\dot{\varphi}\frac{\partial n_{a}}{\partial t}\right) ^{2}=4g^{2}\left(
n_{a}(0)-n_{a}\right) \left( n_{a}-n_{b}(0)\right) .
\end{equation*}

\section{Derivation of Eq. (\protect\ref{s6})}

Taking the time derivative of Eq. (\ref{v1}), using the evolution equation
for the density operator (\ref{q2b}) and the invariance of the trace under
cyclic permutations, we have%
\begin{equation*}
i\hslash \frac{\partial n_{a}}{\partial t}=\text{Tr}\left( [\hat{a}^{\dag }%
\hat{a},\hat{V}]\hat{\rho}(t)\right)
\end{equation*}%
\begin{equation*}
=\hslash g\text{Tr}\left( \left( e^{i\varphi (t)}\hat{a}^{\dag }\hat{b}%
^{\dag }-e^{-i\varphi (t)}\hat{a}\hat{b}\right) \hat{\rho}(t)\right) ,
\end{equation*}%
\begin{equation*}
i\hslash \frac{\partial n_{b}}{\partial t}=\text{Tr}\left( [\hat{b}^{\dag }%
\hat{b},\hat{V}]\hat{\rho}(t)\right)
\end{equation*}%
\begin{equation*}
=\hslash g\text{Tr}\left( \left( e^{i\varphi (t)}\hat{a}^{\dag }\hat{b}%
^{\dag }-e^{-i\varphi (t)}\hat{a}\hat{b}\right) \hat{\rho}(t)\right) ,
\end{equation*}%
or%
\begin{equation*}
-\hslash \frac{\partial }{\partial t}\left( n_{a}+n_{b}\right) =2i\hslash g%
\text{Tr}\left( \left( e^{i\varphi (t)}\hat{a}^{\dag }\hat{b}^{\dag
}-e^{-i\varphi (t)}\hat{a}\hat{b}\right) \hat{\rho}(t)\right) .
\end{equation*}%
Taking the time derivative again, we find%
\begin{equation*}
-\hslash \frac{\partial ^{2}}{\partial t^{2}}\left( n_{a}+n_{b}\right) =
\end{equation*}%
\begin{equation*}
-2\hslash g\dot{\varphi}\text{Tr}\left( \left( e^{i\varphi (t)}\hat{a}^{\dag
}\hat{b}^{\dag }+e^{-i\varphi (t)}\hat{a}\hat{b}\right) \hat{\rho}(t)\right)
\end{equation*}%
\begin{equation*}
+2g\text{Tr}\left( \left[ e^{i\varphi (t)}\hat{a}^{\dag }\hat{b}^{\dag
}-e^{-i\varphi (t)}\hat{a}\hat{b},\hat{V}\right] \hat{\rho}\right) =
\end{equation*}%
\begin{equation*}
-2\hslash g\dot{\varphi}\text{Tr}\left( \left( e^{i\varphi (t)}\hat{a}^{\dag
}\hat{b}^{\dag }+e^{-i\varphi (t)}\hat{a}\hat{b}\right) \hat{\rho}(t)\right)
\end{equation*}%
\begin{equation*}
+2\hslash g^{2}\text{Tr}\left( \left( \left[ \hat{a}^{\dag }\hat{b}^{\dag },%
\hat{a}\hat{b}\right] -\left[ \hat{a}\hat{b},\hat{a}^{\dag }\hat{b}^{\dag }%
\right] \right) \hat{\rho}\right)
\end{equation*}%
\begin{equation*}
=-2\hslash g\dot{\varphi}\text{Tr}\left( \left( e^{i\varphi (t)}\hat{a}%
^{\dag }\hat{b}^{\dag }+e^{-i\varphi (t)}\hat{a}\hat{b}\right) \hat{\rho}%
(t)\right)
\end{equation*}%
\begin{equation*}
-4\hslash g^{2}\text{Tr}\left( \left( 1+\hat{a}^{\dag }\hat{a}+\hat{b}^{\dag
}\hat{b}\right) \hat{\rho}\right)
\end{equation*}%
\begin{equation*}
=-2\dot{\varphi}V(t)-4\hslash g^{2}\left( 1+n_{a}+n_{b}\right) ,
\end{equation*}%
where%
\begin{equation*}
V(t)=\hslash g\text{Tr}\left( \left( e^{i\varphi (t)}\hat{a}^{\dag }\hat{b}%
^{\dag }+e^{-i\varphi (t)}\hat{a}\hat{b}\right) \hat{\rho}(t)\right) .
\end{equation*}

Thus, we obtain differential equation for $n_{a}+n_{b}$ 
\begin{equation*}
\frac{\partial ^{2}}{\partial t^{2}}\left( n_{a}+n_{b}\right) -\frac{2\dot{%
\varphi}}{\hslash }V(t)-4g^{2}\left( 1+n_{a}+n_{b}\right) =0.
\end{equation*}%
Taking the time derivative of $V(t)$, we have%
\begin{equation*}
\frac{2}{\hslash }\frac{\partial V(t)}{\partial t}=-\dot{\varphi}\frac{%
\partial }{\partial t}\left( n_{a}+n_{b}\right) ,
\end{equation*}%
or%
\begin{equation*}
\frac{2}{\hslash }V(t)=-\int_{0}^{t}dt\dot{\varphi}\frac{\partial }{\partial
t}\left( n_{a}+n_{b}\right) .
\end{equation*}%
Therefore,%
\begin{equation*}
\frac{\partial ^{2}\left( n_{a}+n_{b}\right) }{\partial t^{2}}+\dot{\varphi}%
\int_{0}^{t}dt\dot{\varphi}\frac{\partial \left( n_{a}+n_{b}\right) }{%
\partial t}-4g^{2}\left( 1+n_{a}+n_{b}\right) =0.
\end{equation*}%
Multiplying this equation by $\frac{\partial }{\partial t}\left(
n_{a}+n_{b}\right) $ and integrating over time, we obtain%
\begin{equation*}
\left[ \frac{\partial }{\partial t}\left( n_{a}+n_{b}\right) \right] ^{2}+%
\left[ \int_{0}^{t}dt\dot{\varphi}\frac{\partial }{\partial t}\left(
n_{a}+n_{b}\right) \right] ^{2}
\end{equation*}%
\begin{equation*}
-4g^{2}\left( n_{a}+n_{b}\right) ^{2}-8g^{2}\left( n_{a}+n_{b}\right) =const,
\end{equation*}%
or%
\begin{equation*}
\left[ \frac{\partial }{\partial t}\left( n_{a}+n_{b}\right) \right] ^{2}+%
\left[ \int_{0}^{t}dt\dot{\varphi}\frac{\partial }{\partial t}\left(
n_{a}+n_{b}\right) \right] ^{2}
\end{equation*}%
\begin{equation*}
=4g^{2}\left( \left[ n_{a}+n_{b}+1\right] ^{2}-\left[ n_{a}(0)+n_{b}(0)+1%
\right] ^{2}\right) .
\end{equation*}

\end{document}